\documentclass{aa}  
\usepackage{graphicx}
\usepackage[utf8]{inputenc}
\usepackage{natbib}
\usepackage[mathscr]{euscript}
\newcommand\skulad{Sk\'{u}lad\'{o}ttir et al. in prep.}

\begin{document}

   \title{The first carbon-enhanced metal-poor star found in the Sculptor dwarf spheroidal\thanks{Based on observations made with ESO telescopes at the La Silla Paranal observatory under program IDs 291.B-5036 (director's discretionary time) and 089.B-0304(B)}}


   \author{\'{A}. Sk\'{u}lad\'{o}ttir
   		\inst{1}
        \and
        E. Tolstoy\inst{1}
       	\and
       	S. Salvadori\inst{1}
       	\and
       	V. Hill \inst{2}
       	\and
       	M. Pettini \inst{3}  
       	\and
       	M. D. Shetrone \inst{4}  
        \and
       	E. Starkenburg \inst{5} \thanks{CIFAR Global Scholar}       	
          }

   \institute{
              Kapteyn Astronomical Institute, University of Groningen, PO Box 800, 9700AV Groningen, the Netherlands\\
              \email{asa@astro.rug.nl}
		\and
             Laboratoire Lagrange, Universit\'{e} de Nice Sophia Antipolis, CNRS, Observatoire de la C\^{o}te d’Azur, BP 4229, 06304 Nice Cedex 4, France
		\and
          	Institute of Astronomy, Madingley Road, Cambridge CB3 0HA, England
		\and
			The University of Texas at Austin, McDonald Observatory,32 Fowlkes Rd, McDonald Observatory, Tx 79734-3005, USA  
		\and
			Department of Physics and Astronomy, University of Victoria, PO Box 3055, STN CSC, Victoria BC V8W 3P6, Canada
		}
 
  \abstract{The origin of carbon-enhanced metal-poor (CEMP) stars and their possible connection with the chemical elements produced by the first stellar generation is still highly debated. In contrast to the Galactic halo, not many CEMP stars have been found in the dwarf spheroidal galaxies around the Milky Way. Here we present detailed abundances from ESO VLT/UVES high-resolution spectroscopy for ET0097, the first CEMP star found in the Sculptor dwarf spheroidal, which is one of the best studied dwarf galaxies in the Local Group. This star has $\text{[Fe/H]}=-2.03\pm0.10$, $\text{[C/Fe]}=0.51\pm0.10$ and $\text{[N/Fe]}=1.18\pm0.20$, which is the first nitrogen measurement in this galaxy. The traditional definition of CEMP stars is $\text{[C/Fe]}\geq0.70$, but taking into account that this luminous red giant branch star has undergone mixing, it was intrinsically less nitrogen enhanced and more carbon-rich when it was formed, and so it falls under the definition of CEMP stars, as proposed by \citet{Aoki2007} to account for this effect. By making corrections for this mixing, we conclude that the star had $\text{[C/Fe]}\approx0.8$ during its earlier evolutionary stages. Apart from the enhanced C and N abundances, ET0097 shows no peculiarities in other elements lighter than Zn, and no enhancement of the heavier neutron-capture elements (Ba, La, Ce, Nd, Sm, Eu, Dy), making this a CEMP-no star. However, the star does show signs of the weak $r$-process, with an overabundance of the lighter neutron-capture elements (Sr, Y, Zr). To explain the abundance pattern observed in ET0097, we explore the possibility that this star was enriched by primordial stars. In addition to the detailed abundances for ET0097, we present estimates and upper limits for C abundances in 85 other stars in Sculptor derived from CN molecular lines, including 11 stars with $\text{[Fe/H]}\leq-2$. Combining these limits with observations from the literature, the fraction of CEMP-no stars in Sculptor seems to be significantly lower than in the Galactic halo.
}


   \maketitle

\begin{table*}
\caption{UVES Observations of ET0097.}             
\label{table:obs}      
\centering                          
\begin{tabular}{c c c c c c c c c}        
\hline\hline                 
R.A. & Dec. & $v_r$ (km s$^{-1}$) & Obs. date &  Exp.time (min) & Airmass (start) & Airmass (end) & Seeing\\    
\hline                        
$00^\text{h} 59^\text{m} 12^\text{s}$ & $-33^\circ 46'21''$ & 108.97 $\pm$ 0.32 & 2013-08-17 & 47.65 & 1.014 & 1.042 & 0.71\\
&&& 2013-09-01 & 47.65 & 1.129 & 1.261 & 1.04\\ 
&&& 2013-09-04 & 47.65 & 1.179 & 1.080 & 0.79\\ 
&&& 2013-09-04 & 47.65 & 1.068 & 1.023 & 0.85\\ 
\hline                                   
\end{tabular}
\end{table*}

%
\section{Introduction}

The chemical compositions of stellar photospheres provide detailed information about the interstellar medium (ISM) from which the stars were formed. Of particular interest are the very metal-poor (VMP) stars, $\text{[Fe/H]} \leq-2$ (where $\text{[Fe/H]}=\log_{10}(N_{\text{Fe}}/N_\text{H})_\star - \log_{10}(N_{\text{Fe}}/N_\text{H})_\odot$), in the Milky Way environment, which may still preserve imprints of the first generation of stars and the early chemical evolution of the Galaxy and its surroundings. Recent surveys have revealed that a significant fraction of these low-metallicity stars in the Milky Way halo are enhanced in carbon ([C/Fe]~$\geq$~0.7). The cumulative fraction of these carbon-enhanced metal-poor (CEMP) stars in the halo rises from $\sim$20\% for $\text{[Fe/H]}\leq-2.5$ to $\sim$30\% for $\text{[Fe/H]}\leq-3.0$ and up to 75\% for $\text{[Fe/H]}\leq-4.0$ \citep{Lee2013}. Including the recently discovered carbon-enhanced, hyper iron-poor star with $\text{[Fe/H]}<-7.1$ \citep{Keller2014}, the CEMP fraction at the lowest [Fe/H] becomes even higher. 
 
Traditionally, CEMP stars are categorized by their heavy element abundance patterns. Some are enriched in heavy neutron-capture elements built by the slow-process (such as Ba) and are labeled CEMP-$s$ stars, and those that show significant abundances of heavy elements from the rapid-process (such as Eu) are referred to as CEMP-$r$ stars. Those that show enhancements of elements from both processes are labeled CEMP-$s/r$ stars. Finally, CEMP-no stars show no enhancements of the main $r$- or $s$-process elements. 

The main $s$-process happens in low mass (M~$\lesssim$~4~M$_\odot$) asymptotic giant branch (AGB) stars, while the $r$-process requires a high energy, neutron-rich environment, so sites such as supernovae (e.g., \citealt{Travaglio2004}) and neutron star mergers \citep{Tsujimoto2014} have been proposed. The lighter neutron-capture elements (such as Sr, Y and Zr) are created in the main $r$-process, but are overabundant at lower [Fe/H] compared to the heavier $r$-process elements, so an extra source, the weak $r$-process or the weak $s$-process, is needed to explain the observed abundances \citep{Travaglio2004,Francois2007,Frischknecht2012,Cescutti2014}.  

In general, CEMP-$s$ and CEMP-$s/r$ stars have abundance patterns that suggest mass transfer from a companion in the AGB phase \citep{Lucatello2005}. Thus the C, N, O, and heavy elements for these stars do not reflect the ISM from which they were formed. The available data for CEMP-$s$ stars is consistent with $\sim$100\% binary fraction and a maximum period of $\sim$20,000 days \citep{Starkenburg2014}. 

CEMP-no stars, however, are not especially associated with binaries. Though some of them do belong to binary systems, for many of them close binaries that favor mass transfer can be excluded (e.g., \citealt{Starkenburg2014}). These stars are more frequent, and their carbon enhancement becomes more extreme at lower [Fe/H] (e.g., \citealt{Lee2013,Norris2013}), so it is difficult to explain the abundance pattern of such stars with mass transfer from an AGB-companion. Though other scenarios have been discussed, CEMP-no stars are generally believed to have formed out of carbon-enhanced gas clouds, enriched by low- and/or zero-metallicity stars. Thus, CEMP-no stars could provide direct information on the properties of the first generation of stars. 

Among the proposed sources of C-enrichment in CEMP-no stars (see \citealt{Norris2013} for a detailed overview), two are of particular interest: (i) massive rapidly rotating zero-metallicity stars that produce large amounts of C, N, and O due to distinctive internal burning and mixing episodes \citep{Meynet2006}; (ii) faint SNe, associated with the first generations of stars, which experience mixing and fallback, ejecting large amounts of C but small amounts of Fe and the other heavier elements \citep{UmedaNomoto2003,Iwamoto2005,Tominaga2007}. 

Most of the observed CEMP stars have been found in the Galactic halo and a few have been found in the ultra faint (UF) galaxies around the Milky Way. Two stars with very high carbon values ($\text{[C/Fe]}>2$) have been found in the UF galaxies Bootes \citep{Lai2011} and Segue~I \citep{Norris2013}. No star with $\text{[C/Fe]}>2$ has been found in the more luminous (L$_\text{tot}>10^5$~L$_\odot$), more distant dwarf spheroidal (dSph) galaxies, but some CEMP stars with lower carbon abundances have been observed, such as a star in Sextans dSph with $\text{[C/Fe]}\sim 1$ at $\text{[Fe/H]}\sim-3$ \citep{Honda2011}. In particular, no VMP star has been found in the Sculptor dSph with $\text{[C/Fe]}>0.1$ until now, despite extensive searches for low-metallicity stars (e.g., \citealt{Tafelmeyer2010,Starkenburg2013}). 

Sculptor is a well-studied system with a magnitude of $M_V\approx-11.2$ and a distance of $86\pm 5$~kpc \citep{Pietrzynski2008}. It is at high Galactic latitude ($b=-83^\circ$) and has systemic velocity of $v_{\textsl{hel}}~=~+110.6~\pm~0.5$~km/s. The contamination by foreground Galactic stars is not significant, and most of it can be easily distinguished by velocity (e.g., \citealt{Battaglia2008b}). The star formation history shows a peak in star formation $\sim$13~Gyr ago, with a slow decrease, so the galaxy is dominated by an old stellar population (>10~Gyr old), and has not formed any stars for the last  $\sim$6~Gyr \citep{deBoer2012}.

Large spectroscopic surveys of individual stars have been carried out in the central field of Sculptor. Abundances have been measured for Fe, Mg, Ca, Si, and Ti with intermediate-resolution (IR) spectroscopy \citep{Kirby2009} and high-resolution (HR) spectroscopy for $\sim$100 stars (Dwarf Abundances \& Radial-velocities Team (DART) survey, \citealt{Tolstoy2009}; Hill et al. in prep.). Because of the distance to Sculptor, in general only the brightest stars of the galaxy are available for HR spectroscopy. The HR sample is therefore mostly limited to the upper part of the red giant branch (RGB) ($0\lesssim$~$\log g$~$\lesssim2$). None of the large surveys of Sculptor have included measurements of carbon abundances, but several follow-up spectra of low-metallicity stars have been taken, many of them including C measurements  \citep{Tafelmeyer2010,Frebel2010,Kirby2012,Starkenburg2013}. 

In addition, there have been surveys of Carbon-stars and CH-stars in Sculptor (e.g., \citealt{Azzopardi1986}), some of which that have been followed up with IR spectra (e.g., \citealt{Groenewegen2009}). The carbon-enhancement of these stars is believed to come from internal processes or mass transfer, and does not reflect the ISM from which they were formed. Thus they will not be discussed further in this paper.

The star ET0097 is thus the most inherently carbon-rich star in the Sculptor dSph measured to date, with $\text{[Fe/H]}=-2.03\pm0.10$ and $\text{[C/Fe]}=0.51\pm0.10$. It was even more carbon-enhanced in the past, with $\text{[C/Fe]}\approx0.8$, making it the first CEMP observed in Sculptor. A detailed chemical analysis of this star is presented here, from an HR spectrum observed with the VLT/UVES telescope at the European Southern Observatory (ESO). In addition, carbon abundance estimates and upper limits for 85 other stars are presented, derived from CN molecular lines in the wavelength range 9100-9250 \text{\AA}, from VLT/FLAMES spectra.

%
\section{Observations and data reduction}

From the DART survey \citep{Tolstoy2006}, detailed abundance measurements are known for $\sim$100 stars, spread over a $25^\prime$ diameter field of view in the Sculptor dSph (Hill et al. in prep.; \citealt{Tolstoy2009}). As a part of this project, ESO VLT/FLAMES/GIRAFFE HR spectroscopy was carried out over the wavelength range $\sim$9100-9300~\text{\AA}, to measure S abundances in Sculptor (Sk\'{u}lad\'{o}ttir et al. in prep.). In most of these spectra CN molecular lines were observed, with the exception of the most metal-poor stars ($\text{[Fe/H]}\lesssim-2.2$). This CN molecular band was exceptionally strong in the star ET0097 and to follow up that observation, an HR spectrum over a long wavelength range was taken for the star, using ESO VLT/UVES. 

UVES is a dichroic HR optical spectrograph at the VLT \citep{Dekker2000}, where the light beam from the telescope can be split into two arms, the Ultra Violet to the Blue arm and the Visual to the Red arm. The observations were taken in August and September of 2013, using a 1.2'' slit, with a resolution of 40,000 in the blue, and 35,000 in the red. The observational details are listed in Table~\ref{table:obs}. 

The details regarding the observations and data reduction for the VLT/FLAMES data, along with the S measurements will be presented in an upcoming paper, Sk\'ulad\'ottir et al. in prep.

\begin{table}\label{tab:SNR}
\caption{Signal-to-noise ratios of the different parts of the final coadded spectrum.}             
\centering                          
\begin{tabular}{c c}        
\hline\hline                 
Wavelength range (\text{\AA}) & S/N\\    
\hline                        
3770-4980	&	30\\ 
5760-7510	&	50\\ 
7660-9450	&	50\\ 
\hline                                   
\end{tabular}
\end{table}

\subsection{Data reduction}

The ESO VLT/UVES spectrum was reduced, extracted, wavelength calibrated, and sky-subtracted using the UVES pipeline provided by ESO \citep{Freudling2013}. The reduced spectra were corrected for telluric absorption using spectra of a blue horizontal branch star, taken the same nights as the observations. The spectra taken at different times all showed comparable counts, so they were combined using a median value of the four spectra. The usable wavelength range and their relative signal-to-noise (S/N) ratios are listed in Table \ref{tab:SNR}. The S/N ratios were evaluated as the mean value over the standard deviation of the continuum in line-free regions.

\subsection{Continuum normalization}

In the red part of the spectrum ($\sim$5800-9400~\text{\AA}), the entire wavelength range was covered with CN molecular lines, most of them weak, but some stronger. To find proper continuum points for the spectrum, a synthetic spectrum was made, using rough estimates of the oxygen, carbon, and nitrogen abundances. An iterative comparison with the normalized observed spectrum was then used to find a better synthetic spectrum, which was used for a better determination of continuum points. This process was iterated until the result was stable.

A similar approach was used for the reddest part of the blue spectrum ($\sim$4500-5000 \AA) that is covered in relatively weak CH and C$_2$ molecular lines. For the bluer part of the spectrum, true continuum points became rarer, and a continuum could only be estimated from points close to the continuum. At the bluest part of the spectrum, the B-X band of CN at 3888~\AA$ $ is extremely strong, and wipes out all continuum points making normalization in the region very uncertain, so the bluest part of the spectrum ($\lesssim3900$~\AA) could not be used in the abundance analysis.

Where enough continuum points were available, the spectrum was also renormalized around each line being measured by a constant factor for better accuracy, but this change in the height of the continuum was minimal, rarely more than 1-2\%.

\begin{table}
\caption{Photometry of ET0097.}          
\label{table:photo}      
\centering                          
\begin{tabular}{c c c c c}        
\hline\hline                 
Colour & mag. & err & E(V-X) & $T_\textsl{eff}$ (K) \\    
\hline                        
   V   & 17.255 & 0.002 &  & \\
   V-I &  1.246 & 0.005 & 0.023 & 4382 \\ 
   V-J &  2.125 & 0.004 & 0.041 & 4393 \\
   V-K &  2.954 & 0.005 & 0.050 & 4378 \\
\hline 
\end{tabular}
\end{table}

%
\section{Stellar parameters}

The photometry for ET0097 comes from deep wide field imaging in the V and I bands \citep{deBoer2011} and the infrared photometry, bands J and K, come from VISTA survey observations, see Table~\ref{table:photo}. Although photometry is also available for the B band, it was not used, since in this carbon-rich star it is affected by the strong CH molecular band in the region.

The effective temperature of the star, $T_\textsl{eff}$, was determined from the photometry following the recipe from \citet{RamirezMelendez2005} for giants with metallicity $\text{[Fe/H]}=-2.02$ and assuming a reddening correction, E(V-X), in the direction of the Sculptor dSph as evaluated by \citet{Schlegel1998}, see Table \ref{table:photo}. The result, $T_\textsl{eff}~=~4383~\pm~35$~K, is 83~K higher than the temperature adopted for this star in Hill et al. in prep., but agrees within the error bars. Included in the estimated errors are the $\sigma(T_\textsl{eff})$ provided by \citet{RamirezMelendez2005} for each color used, the photometric errors and the (minor) effect of the observational errors of [Fe/H].

The surface gravity for the star is obtained using the standard relation:
\begin{equation}
\log g_{\star}=\log g_{\odot}+\log{\frac{\text{M}_{\star}}{\text{M}_{\odot}}}+  4\log{ \frac{T_{\textsl{eff,}\star}}{T_{\textsl{eff,}\odot}} }+0.4(M_{\textsl{bol,}\star}-M_{\textsl{bol,}\odot})
\end{equation}
which yields the surface gravity $\log g_\star = 0.75\pm0.13$.

The absolute bolometric magnitude for the star, $M_{\textsl{bol,}\star}=-3.01\pm0.08$, is calculated using a calibration for the V-band magnitude \citep{Alonso1999} and a distance modulus of $(m-M)_0~=~19.68\pm0.08$, from \citet{Pietrzynski2008}, which dominates the error on $M_{\textsl{bol,}\star}$, as the photometric errors are negligible. The mass of the star is assumed to be $\text{M}_\star~=~0.8\pm0.2 $~M$_\odot$. The solar values used are the following: $\log~g_\odot=4.44$, $T_{\textsl{eff,}\odot}~=~5790$~K and $M_{\textsl{bol,}\odot}~=~4.72$.

The microturbulence velocity, $v_t$, was spectroscopically derived by making sure the abundance measurements for Fe~I do not show a trend with the equivalent width, $\log($EW$/\lambda)$. The result is $v_t~=~2.25~\pm~0.20$~km/s.

No significant slope was found between the iron abundances of individual lines and their relevant wavelength or their excitation potential, confirming the validity of the determined stellar parameters.

\begin{longtab}
\begin{longtable}{l r r r r r l}
\caption{Linelist and abundance measurements for individual lines. Errors are included for elements with fewer than five measured lines. A line where the abundance is marked with an hyphen, is fitted together with the previous line/lines.}    \\                       
\hline   
X$_i$	&	$\lambda$	&	$\chi_{ex}$	&	$\log (gf)$	& 	$\log\epsilon(X_i)$	&	$\delta_\textsl{noise,i}$ & Comment	\\
\hline   
\endfirsthead
\caption{continued.}\\
 \hline                 
X$_i$	&	$\lambda$	&	$\chi_{ex}$	&	$\log (gf)$	& 	$\log\epsilon(X_i)$	&	$\delta_\textsl{noise,i}$ & Comment	\\
\hline   
\endhead
Li I	&$	6707.76	$&$	0.000	$&$	-0.009	$&$	<0.17	$&$	$-$	$&	blended	\\
Li I	&$	6707.91	$&$	0.000	$&$	-0.309	$&$	$-$	$&$	$-$	$&		\\
\hline													
O I	&$	6300.30	$&$	0.000	$&$	-9.819	$&$	7.30	$&$	0.14	$&		\\
O I	&$	6363.78	$&$	0.020	$&$	-10.303	$&$	7.46	$&$	0.30	$&		\\
\hline													
Na I	&$	5889.95	$&$	0.000	$&$	0.117	$&$	3.90	$&$	0.26	$&		\\
Na I	&$	5895.92	$&$	0.000	$&$	-0.184	$&$	3.68	$&$	0.20	$&		\\
Na I	&$	8183.26	$&$	2.102	$&$	0.230	$&$	3.94	$&$	0.28	$&		\\
\hline													
Mg I	&$	4571.10	$&$	0.000	$&$	-5.691	$&$	6.02	$&$	$-$	$&		\\
Mg I	&$	4702.99	$&$	4.346	$&$	-0.666	$&$	5.80	$&$	$-$	$&		\\
Mg I	&$	5711.09	$&$	4.346	$&$	-1.833	$&$	5.96	$&$	$-$	$&		\\
Mg I	&$	8736.01	$&$	5.946	$&$	-3.210	$&$	6.04	$&$	$-$	$&		\\
Mg I	&$	8736.01	$&$	5.946	$&$	-1.930	$&$	$-$	$&$	$-$	$&		\\
Mg I	&$	8736.02	$&$	5.946	$&$	-3.300	$&$	$-$	$&$	$-$	$&		\\
Mg I	&$	8736.02	$&$	5.946	$&$	-0.690	$&$	$-$	$&$	$-$	$&		\\
Mg I	&$	8736.02	$&$	5.946	$&$	-1.970	$&$	$-$	$&$	$-$	$&		\\
Mg I	&$	8736.03	$&$	5.946	$&$	-1.020	$&$	$-$	$&$	$-$	$&		\\
Mg I	&$	8806.76	$&$	4.346	$&$	-0.137	$&$	6.02	$&$	$-$	$&		\\
\hline													
Al I	&$	3961.52	$&$	0.014	$&$	-0.323	$&$	3.62	$&$	0.80	$&	very blended	\\
\hline													
Si I	&$	5708.40	$&$	4.954	$&$	-1.470	$&$	5.86	$&$	$-$	$&		\\
Si I	&$	5948.54	$&$	5.082	$&$	-1.230	$&$	5.80	$&$	$-$	$&	blended	\\
Si I	&$	7034.90	$&$	5.871	$&$	-0.880	$&$	6.12	$&$	$-$	$&		\\
Si I	&$	7275.26	$&$	6.206	$&$	-7.048	$&$	5.90	$&$	$-$	$&	blended	\\
Si I	&$	7275.26	$&$	6.206	$&$	-8.389	$&$	$-$	$&$	$-$	$&		\\
Si I	&$	7275.30	$&$	5.616	$&$	-0.847	$&$	$-$	$&$	$-$	$&		\\
Si I	&$	7409.08	$&$	5.616	$&$	-0.880	$&$	5.78	$&$	$-$	$&		\\
Si I	&$	7409.15	$&$	5.964	$&$	-1.566	$&$	$-$	$&$	$-$	$&		\\
Si I	&$	7423.50	$&$	5.619	$&$	-0.175	$&$	5.42	$&$	$-$	$&		\\
Si I	&$	8752.01	$&$	5.871	$&$	0.079	$&$	5.50	$&$	$-$	$&		\\
\hline													
S I	&$	9212.86	$&$	6.525	$&$	0.420	$&$	5.42	$&$	0.32	$&		\\
S I	&$	9228.09	$&$	6.525	$&$	0.260	$&$	5.46	$&$	0.26	$&		\\
\hline													
K I	&$	7664.91	$&$	0.000	$&$	0.130	$&$	3.46	$&$	0.10	$&		\\
K I	&$	7698.97	$&$	0.000	$&$	-0.170	$&$	3.44	$&$	0.06	$&		\\
\hline													
Ca I	&$	5857.45	$&$	2.933	$&$	0.240	$&$	4.40	$&$	$-$	$&		\\
Ca I	&$	6102.44	$&$	2.523	$&$	-2.805	$&$	4.46	$&$	$-$	$&		\\
Ca I	&$	6102.72	$&$	1.879	$&$	-0.793	$&$	$-$	$&$	$-$	$&		\\
Ca I	&$	6122.22	$&$	1.886	$&$	-0.316	$&$	4.52	$&$	$-$	$&		\\
Ca I	&$	6161.30	$&$	2.523	$&$	-1.266	$&$	4.50	$&$	$-$	$&		\\
Ca I	&$	6162.17	$&$	1.899	$&$	-0.090	$&$	4.50	$&$	$-$	$&		\\
Ca I	&$	6166.44	$&$	2.521	$&$	-1.142	$&$	4.56	$&$	$-$	$&		\\
Ca I	&$	6169.04	$&$	2.523	$&$	-0.797	$&$	4.60	$&$	$-$	$&		\\
Ca I	&$	6169.56	$&$	2.526	$&$	-0.478	$&$	4.48	$&$	$-$	$&		\\
Ca I	&$	6439.08	$&$	2.526	$&$	0.390	$&$	4.44	$&$	$-$	$&		\\
Ca I	&$	6439.17	$&$	5.490	$&$	-3.709	$&$	$-$	$&$	$-$	$&		\\
Ca I	&$	6439.24	$&$	5.832	$&$	-4.094	$&$	$-$	$&$	$-$	$&		\\
Ca I	&$	6449.81	$&$	2.521	$&$	-0.502	$&$	4.48	$&$	$-$	$&		\\
Ca I	&$	6455.60	$&$	2.523	$&$	-1.340	$&$	4.56	$&$	$-$	$&	blended	\\
Ca I	&$	6462.57	$&$	2.523	$&$	0.262	$&$	4.34	$&$	$-$	$&	blended	\\
Ca I	&$	6471.66	$&$	2.526	$&$	-0.686	$&$	4.46	$&$	$-$	$&		\\
Ca I	&$	6493.78	$&$	2.521	$&$	-0.109	$&$	4.48	$&$	$-$	$&		\\
Ca I	&$	6499.65	$&$	2.523	$&$	-0.818	$&$	4.50	$&$	$-$	$&		\\
Ca I	&$	6572.78	$&$	0.000	$&$	-4.240	$&$	4.44	$&$	$-$	$&		\\
Ca I	&$	6717.68	$&$	2.709	$&$	-0.524	$&$	4.60	$&$	$-$	$&		\\
Ca I	&$	6717.69	$&$	5.883	$&$	-7.108	$&$	$-$	$&$	$-$	$&		\\
Ca I	&$	7148.15	$&$	2.709	$&$	0.137	$&$	4.58	$&$	$-$	$&		\\
Ca I	&$	7202.20	$&$	2.709	$&$	-0.262	$&$	4.44	$&$	$-$	$&		\\
Ca I	&$	7202.56	$&$	6.021	$&$	-4.660	$&$	$-$	$&$	$-$	$&		\\
Ca I	&$	7326.15	$&$	2.933	$&$	-0.208	$&$	4.42	$&$	$-$	$&		\\
Ca I	&$	7326.48	$&$	6.023	$&$	-5.128	$&$	$-$	$&$	$-$	$&		\\
Ca I	&$	7326.48	$&$	6.007	$&$	-4.670	$&$	$-$	$&$	$-$	$&		\\
\hline													
Sc II	&$	4246.82	$&$	0.315	$&$	0.242	$&$	1.14	$&$	$-$	$&	blended	\\
Sc II	&$	4294.77	$&$	0.605	$&$	-1.391	$&$	0.90	$&$	$-$	$&	very blended	\\
Sc I	&$	4320.62	$&$	2.109	$&$	-1.920	$&$	1.00	$&$	$-$	$&	blended	\\
Sc II	&$	4320.73	$&$	0.605	$&$	-0.252	$&$	$-$	$&$	$-$	$&		\\
Sc I	&$	4415.48	$&$	3.083	$&$	-3.393	$&$	1.24	$&$	$-$	$&	blended	\\
Sc II	&$	4415.56	$&$	0.595	$&$	-0.668	$&$	$-$	$&$	$-$	$&		\\
Sc I	&$	4431.23	$&$	1.851	$&$	-6.387	$&$	1.16	$&$	$-$	$&		\\
Sc II	&$	4431.35	$&$	0.605	$&$	-1.969	$&$	$-$	$&$	$-$	$&		\\
Sc I	&$	4431.52	$&$	3.083	$&$	-2.584	$&$	$-$	$&$	$-$	$&		\\
Sc II	&$	4670.41	$&$	1.357	$&$	-0.576	$&$	0.92	$&$	$-$	$&	very blended	\\
Sc I	&$	4670.52	$&$	3.172	$&$	-2.584	$&$	$-$	$&$	$-$	$&		\\
Sc II	&$	6245.64	$&$	1.507	$&$	-1.030	$&$	1.24	$&$	$-$	$&	blended	\\
Sc I	&$	6279.57	$&$	3.607	$&$	-1.673	$&$	1.18	$&$	$-$	$&		\\
Sc II	&$	6279.75	$&$	1.500	$&$	-1.265	$&$	$-$	$&$	$-$	$&		\\
Sc II	&$	6604.60	$&$	1.357	$&$	-1.309	$&$	1.16	$&$	$-$	$&		\\
\hline												
Ti II	&$	4493.51	$&$	1.080	$&$	-3.020	$&$	3.12	$&$	$-$	$&		\\
Ti II	&$	4501.27	$&$	1.116	$&$	-0.770	$&$	2.78	$&$	$-$	$&		\\
Ti I	&$	4562.63	$&$	0.021	$&$	-2.656	$&$	3.00	$&$	$-$	$&		\\
Ti I	&$	4563.42	$&$	2.427	$&$	-0.681	$&$	2.92	$&$	$-$	$&		\\
Ti II	&$	4563.76	$&$	1.221	$&$	-0.690	$&$	$-$	$&$	$-$	$&		\\
Ti II	&$	4568.31	$&$	1.224	$&$	-2.940	$&$	3.08	$&$	$-$	$&		\\
Ti II	&$	4583.41	$&$	1.165	$&$	-2.920	$&$	3.18	$&$	$-$	$&		\\
Ti II	&$	4609.26	$&$	1.180	$&$	-3.430	$&$	3.26	$&$	$-$	$&		\\
Ti I	&$	4609.34	$&$	3.319	$&$	-2.072	$&$	$-$	$&$	$-$	$&		\\
Ti I	&$	4617.27	$&$	1.749	$&$	0.389	$&$	2.96	$&$	$-$	$&		\\
Ti I	&$	4708.42	$&$	3.199	$&$	-3.799	$&$	3.18	$&$	$-$	$&	blended	\\
Ti I	&$	4708.43	$&$	1.873	$&$	-6.885	$&$	$-$	$&$	$-$	$&		\\
Ti II	&$	4708.66	$&$	1.237	$&$	-2.340	$&$	$-$	$&$	$-$	$&		\\
Ti I	&$	4759.14	$&$	2.778	$&$	-2.543	$&$	3.12	$&$	$-$	$&		\\
Ti I	&$	4759.27	$&$	2.256	$&$	0.514	$&$	$-$	$&$	$-$	$&		\\
Ti II	&$	4764.52	$&$	1.237	$&$	-2.950	$&$	3.50	$&$	$-$	$&		\\
Ti I	&$	4792.25	$&$	0.813	$&$	-3.468	$&$	3.28	$&$	$-$	$&	blended	\\
Ti II	&$	4792.43	$&$	1.237	$&$	-3.328	$&$	$-$	$&$	$-$	$&		\\
Ti I	&$	4792.48	$&$	2.334	$&$	-0.300	$&$	$-$	$&$	$-$	$&		\\
Ti II	&$	4805.08	$&$	2.061	$&$	-0.960	$&$	3.14	$&$	$-$	$&		\\
Ti I	&$	4805.42	$&$	2.345	$&$	0.150	$&$	$-$	$&$	$-$	$&		\\
Ti I	&$	4805.44	$&$	3.062	$&$	-3.409	$&$	$-$	$&$	$-$	$&		\\
Ti I	&$	4840.87	$&$	0.900	$&$	-0.509	$&$	3.04	$&$	$-$	$&		\\
Ti II	&$	4865.61	$&$	1.116	$&$	-2.790	$&$	3.32	$&$	$-$	$&		\\
Ti I	&$	4865.78	$&$	2.578	$&$	-0.398	$&$	$-$	$&$	$-$	$&		\\
Ti III	&$	4874.00	$&$	18.252	$&$	0.560	$&$	3.12	$&$	$-$	$&		\\
Ti II	&$	4874.01	$&$	3.095	$&$	-0.800	$&$	$-$	$&$	$-$	$&		\\
Ti I	&$	4885.08	$&$	1.887	$&$	0.358	$&$	2.96	$&$	$-$	$&	blended	\\
Ti I	&$	4885.20	$&$	2.677	$&$	-1.681	$&$	$-$	$&$	$-$	$&		\\
Ti I	&$	4913.46	$&$	2.506	$&$	-3.499	$&$	3.18	$&$	$-$	$&		\\
Ti I	&$	4913.61	$&$	1.873	$&$	0.160	$&$	$-$	$&$	$-$	$&		\\
Ti I	&$	4981.73	$&$	0.848	$&$	0.504	$&$	2.88	$&$	$-$	$&		\\
Ti I	&$	4981.90	$&$	2.427	$&$	-3.637	$&$	$-$	$&$	$-$	$&		\\
Ti I	&$	5866.23	$&$	3.176	$&$	-3.522	$&$	3.12	$&$	$-$	$&		\\
Ti I	&$	5866.37	$&$	3.305	$&$	-0.186	$&$	$-$	$&$	$-$	$&		\\
Ti I	&$	5866.45	$&$	1.067	$&$	-0.840	$&$	$-$	$&$	$-$	$&		\\
Ti II	&$	5866.66	$&$	8.089	$&$	-0.620	$&$	$-$	$&$	$-$	$&		\\
Ti II	&$	5899.03	$&$	8.082	$&$	-2.325	$&$	3.16	$&$	$-$	$&		\\
Ti I	&$	5899.29	$&$	1.053	$&$	-1.154	$&$	$-$	$&$	$-$	$&		\\
Ti I	&$	5899.50	$&$	3.351	$&$	-2.307	$&$	$-$	$&$	$-$	$&		\\
Ti I	&$	5921.92	$&$	3.691	$&$	-2.186	$&$	3.22	$&$	$-$	$&		\\
Ti II	&$	5921.94	$&$	8.056	$&$	-0.024	$&$	$-$	$&$	$-$	$&		\\
Ti I	&$	5922.11	$&$	1.046	$&$	-1.466	$&$	$-$	$&$	$-$	$&		\\
Ti I	&$	5922.14	$&$	3.113	$&$	-1.602	$&$	$-$	$&$	$-$	$&		\\
Ti II	&$	5952.98	$&$	8.071	$&$	-0.063	$&$	3.00	$&$	$-$	$&		\\
Ti I	&$	5953.11	$&$	3.090	$&$	-3.612	$&$	$-$	$&$	$-$	$&		\\
Ti I	&$	5953.16	$&$	1.887	$&$	-0.329	$&$	$-$	$&$	$-$	$&		\\
Ti I	&$	5965.32	$&$	3.409	$&$	-3.412	$&$	3.14	$&$	$-$	$&	blended	\\
Ti II	&$	5965.81	$&$	8.089	$&$	-1.623	$&$	$-$	$&$	$-$	$&		\\
Ti I	&$	5965.83	$&$	1.879	$&$	-0.409	$&$	$-$	$&$	$-$	$&		\\
Ti II	&$	5978.41	$&$	8.093	$&$	-0.535	$&$	3.14	$&$	$-$	$&		\\
Ti I	&$	5978.54	$&$	1.873	$&$	-0.496	$&$	$-$	$&$	$-$	$&		\\
Ti II	&$	6126.00	$&$	8.097	$&$	-1.650	$&$	3.22	$&$	$-$	$&		\\
Ti I	&$	6126.22	$&$	1.067	$&$	-1.425	$&$	$-$	$&$	$-$	$&		\\
Ti I	&$	6126.27	$&$	3.154	$&$	-3.025	$&$	$-$	$&$	$-$	$&		\\
Ti I	&$	6257.80	$&$	0.000	$&$	-4.297	$&$	3.04	$&$	$-$	$&		\\
Ti I	&$	6258.10	$&$	1.443	$&$	-0.355	$&$	$-$	$&$	$-$	$&		\\
Ti I	&$	6258.71	$&$	1.460	$&$	-0.240	$&$	3.14	$&$	$-$	$&		\\
Ti I	&$	6261.10	$&$	1.430	$&$	-0.479	$&$	3.16	$&$	$-$	$&		\\
Ti I	&$	6261.12	$&$	3.128	$&$	-1.798	$&$	$-$	$&$	$-$	$&		\\
Ti I	&$	6261.28	$&$	3.323	$&$	-3.656	$&$	$-$	$&$	$-$	$&		\\
Ti II	&$	6491.56	$&$	2.061	$&$	-1.793	$&$	2.98	$&$	$-$	$&		\\
Ti I	&$	6556.06	$&$	1.460	$&$	-1.074	$&$	3.38	$&$	$-$	$&		\\
Ti II	&$	6559.59	$&$	2.048	$&$	-2.019	$&$	3.04	$&$	$-$	$&		\\
Ti I	&$	7209.44	$&$	1.460	$&$	-0.500	$&$	3.14	$&$	$-$	$&	blended	\\
Ti I	&$	7209.62	$&$	3.424	$&$	-1.599	$&$	$-$	$&$	$-$	$&		\\
Ti I	&$	7244.85	$&$	1.443	$&$	-0.810	$&$	3.24	$&$	$-$	$&	blended	\\
Ti I	&$	7251.71	$&$	1.430	$&$	-0.770	$&$	3.02	$&$	$-$	$&	blended	\\
\hline													
V II	&$	4002.80	$&$	1.555	$&$	-5.601	$&$	2.14	$&$	$-$	$&		\\
V II	&$	4002.94	$&$	1.428	$&$	-1.447	$&$	$-$	$&$	$-$	$&		\\
V I	&$	4003.01	$&$	2.332	$&$	-3.063	$&$	$-$	$&$	$-$	$&		\\
V I	&$	4003.12	$&$	2.359	$&$	-4.625	$&$	$-$	$&$	$-$	$&		\\
V I	&$	4003.13	$&$	2.505	$&$	-9.645	$&$	$-$	$&$	$-$	$&		\\
V I	&$	4003.17	$&$	2.578	$&$	-1.161	$&$	$-$	$&$	$-$	$&		\\
V I	&$	4586.23	$&$	2.578	$&$	-6.028	$&$	1.88	$&$	$-$	$&		\\
V I	&$	4586.37	$&$	0.040	$&$	-0.790	$&$	$-$	$&$	$-$	$&		\\
V I	&$	4586.44	$&$	1.376	$&$	-6.348	$&$	$-$	$&$	$-$	$&		\\
V I	&$	4586.55	$&$	2.505	$&$	-3.017	$&$	$-$	$&$	$-$	$&		\\
V I	&$	4827.46	$&$	0.040	$&$	-1.478	$&$	1.94	$&$	$-$	$&		\\
V II	&$	4827.47	$&$	8.645	$&$	-3.405	$&$	$-$	$&$	$-$	$&		\\
V I	&$	4827.54	$&$	2.040	$&$	-5.969	$&$	$-$	$&$	$-$	$&		\\
V I	&$	4827.69	$&$	2.878	$&$	-3.382	$&$	$-$	$&$	$-$	$&		\\
V I	&$	4831.65	$&$	0.017	$&$	-1.380	$&$	1.96	$&$	$-$	$&		\\
V I	&$	4831.77	$&$	1.955	$&$	-3.283	$&$	$-$	$&$	$-$	$&		\\
V I	&$	4864.73	$&$	0.017	$&$	-0.960	$&$	1.78	$&$	$-$	$&		\\
V I	&$	4864.83	$&$	1.183	$&$	-1.240	$&$	$-$	$&$	$-$	$&		\\
V II	&$	4864.87	$&$	6.857	$&$	-2.457	$&$	$-$	$&$	$-$	$&		\\
V II	&$	4875.22	$&$	6.547	$&$	-4.285	$&$	1.98	$&$	$-$	$&		\\
V I	&$	4875.42	$&$	1.351	$&$	-4.161	$&$	$-$	$&$	$-$	$&		\\
V I	&$	4875.49	$&$	0.040	$&$	-0.810	$&$	$-$	$&$	$-$	$&		\\
V II	&$	4875.50	$&$	5.468	$&$	-1.533	$&$	$-$	$&$	$-$	$&		\\
V II	&$	4875.62	$&$	4.005	$&$	-3.268	$&$	$-$	$&$	$-$	$&		\\
V I	&$	5703.58	$&$	1.051	$&$	-0.211	$&$	2.20	$&$	$-$	$&		\\
V II	&$	5703.65	$&$	3.973	$&$	-4.727	$&$	$-$	$&$	$-$	$&		\\
V II	&$	5706.77	$&$	6.901	$&$	-3.971	$&$	2.18	$&$	$-$	$&		\\
V II	&$	5706.86	$&$	9.031	$&$	-1.107	$&$	$-$	$&$	$-$	$&		\\
V I	&$	5706.98	$&$	1.043	$&$	-0.454	$&$	$-$	$&$	$-$	$&		\\
V I	&$	6039.72	$&$	1.064	$&$	-0.650	$&$	2.10	$&$	$-$	$&		\\
V I	&$	6039.86	$&$	3.517	$&$	-2.920	$&$	$-$	$&$	$-$	$&		\\
V I	&$	6039.86	$&$	2.555	$&$	-4.421	$&$	$-$	$&$	$-$	$&		\\
V I	&$	6090.21	$&$	1.081	$&$	-0.062	$&$	1.98	$&$	$-$	$&		\\
V II	&$	6090.47	$&$	3.799	$&$	-6.833	$&$	$-$	$&$	$-$	$&		\\
V I	&$	6090.54	$&$	1.064	$&$	-2.600	$&$	$-$	$&$	$-$	$&		\\
\hline													
Cr II	&$	4545.94	$&$	8.348	$&$	-2.289	$&$	2.96	$&$	$-$	$&		\\
Cr I	&$	4545.95	$&$	0.941	$&$	-1.370	$&$	$-$	$&$	$-$	$&		\\
Cr I	&$	4546.02	$&$	3.551	$&$	-3.528	$&$	$-$	$&$	$-$	$&		\\
Cr II	&$	4546.03	$&$	6.805	$&$	-3.416	$&$	$-$	$&$	$-$	$&		\\
Cr II	&$	4588.20	$&$	4.071	$&$	-0.845	$&$	3.84	$&$	$-$	$&		\\
Cr I	&$	4588.24	$&$	4.389	$&$	-3.887	$&$	$-$	$&$	$-$	$&		\\
Cr II	&$	4588.40	$&$	3.104	$&$	-4.542	$&$	$-$	$&$	$-$	$&		\\
Cr I	&$	4591.13	$&$	4.402	$&$	-3.455	$&$	3.34	$&$	$-$	$&		\\
Cr I	&$	4591.39	$&$	0.968	$&$	-1.740	$&$	$-$	$&$	$-$	$&		\\
Cr II	&$	4591.42	$&$	10.599	$&$	-4.794	$&$	$-$	$&$	$-$	$&		\\
Cr I	&$	4591.46	$&$	4.440	$&$	-1.925	$&$	$-$	$&$	$-$	$&		\\
Cr I	&$	4591.48	$&$	3.422	$&$	-1.888	$&$	$-$	$&$	$-$	$&		\\
Cr I	&$	4591.60	$&$	4.490	$&$	-3.992	$&$	$-$	$&$	$-$	$&		\\
Cr I	&$	4616.12	$&$	0.983	$&$	-1.190	$&$	3.18	$&$	$-$	$&		\\
Cr II	&$	4616.24	$&$	5.670	$&$	-2.346	$&$	$-$	$&$	$-$	$&		\\
Cr I	&$	4625.92	$&$	3.850	$&$	-0.310	$&$	3.26	$&$	$-$	$&		\\
Cr II	&$	4625.94	$&$	11.677	$&$	-0.760	$&$	$-$	$&$	$-$	$&		\\
Cr I	&$	4626.02	$&$	4.532	$&$	-0.960	$&$	$-$	$&$	$-$	$&		\\
Cr II	&$	4626.09	$&$	11.788	$&$	-4.123	$&$	$-$	$&$	$-$	$&		\\
Cr I	&$	4626.17	$&$	0.968	$&$	-1.320	$&$	$-$	$&$	$-$	$&		\\
Cr I	&$	4634.00	$&$	3.551	$&$	-1.808	$&$	3.70	$&$	$-$	$&	blended	\\
Cr II	&$	4634.07	$&$	4.072	$&$	-1.236	$&$	$-$	$&$	$-$	$&		\\
Cr II	&$	4651.13	$&$	11.622	$&$	-1.756	$&$	3.44	$&$	$-$	$&		\\
Cr II	&$	4651.23	$&$	10.243	$&$	-5.278	$&$	$-$	$&$	$-$	$&		\\
Cr I	&$	4651.28	$&$	0.983	$&$	-1.460	$&$	$-$	$&$	$-$	$&		\\
Cr II	&$	4651.40	$&$	11.711	$&$	-3.420	$&$	$-$	$&$	$-$	$&		\\
Cr II	&$	4652.04	$&$	10.476	$&$	-1.599	$&$	3.32	$&$	$-$	$&		\\
Cr I	&$	4652.16	$&$	1.004	$&$	-1.030	$&$	$-$	$&$	$-$	$&		\\
Cr II	&$	4652.27	$&$	5.871	$&$	-4.565	$&$	$-$	$&$	$-$	$&		\\
Cr II	&$	4755.95	$&$	7.899	$&$	-5.302	$&$	3.84	$&$	$-$	$&		\\
Cr I	&$	4756.05	$&$	2.987	$&$	-2.912	$&$	$-$	$&$	$-$	$&		\\
Cr I	&$	4756.11	$&$	3.104	$&$	0.090	$&$	$-$	$&$	$-$	$&		\\
Cr I	&$	4756.16	$&$	4.106	$&$	-3.511	$&$	$-$	$&$	$-$	$&		\\
Cr I	&$	4756.31	$&$	4.411	$&$	-4.185	$&$	$-$	$&$	$-$	$&		\\
Cr II	&$	4829.18	$&$	10.476	$&$	-2.906	$&$	3.96	$&$	$-$	$&		\\
Cr II	&$	4829.22	$&$	10.798	$&$	-2.554	$&$	$-$	$&$	$-$	$&		\\
Cr I	&$	4829.22	$&$	3.369	$&$	-2.423	$&$	$-$	$&$	$-$	$&		\\
Cr I	&$	4829.31	$&$	2.545	$&$	-1.604	$&$	$-$	$&$	$-$	$&		\\
Cr I	&$	4829.37	$&$	2.545	$&$	-0.810	$&$	$-$	$&$	$-$	$&		\\
Cr I	&$	4848.06	$&$	5.211	$&$	-1.253	$&$	3.70	$&$	$-$	$&		\\
Cr II	&$	4848.23	$&$	3.864	$&$	-1.280	$&$	$-$	$&$	$-$	$&		\\
Cr I	&$	4876.40	$&$	4.096	$&$	-2.973	$&$	3.76	$&$	$-$	$&		\\
Cr II	&$	4876.40	$&$	3.854	$&$	-1.580	$&$	$-$	$&$	$-$	$&		\\
Cr II	&$	4876.47	$&$	3.864	$&$	-2.093	$&$	$-$	$&$	$-$	$&		\\
Cr II	&$	4876.67	$&$	6.686	$&$	-2.966	$&$	$-$	$&$	$-$	$&		\\
Cr II	&$	4876.69	$&$	8.363	$&$	-6.334	$&$	$-$	$&$	$-$	$&		\\
Cr I	&$	4922.16	$&$	3.435	$&$	-3.337	$&$	3.50	$&$	$-$	$&		\\
Cr II	&$	4922.22	$&$	10.843	$&$	-2.464	$&$	$-$	$&$	$-$	$&		\\
Cr I	&$	4922.27	$&$	3.104	$&$	0.270	$&$	$-$	$&$	$-$	$&		\\
Cr II	&$	4922.36	$&$	10.454	$&$	-2.474	$&$	$-$	$&$	$-$	$&		\\
Cr I	&$	4922.54	$&$	3.011	$&$	-2.173	$&$	$-$	$&$	$-$	$&		\\
Cr I	&$	4942.50	$&$	0.941	$&$	-2.294	$&$	3.58	$&$	$-$	$&		\\
Cr I	&$	4942.75	$&$	3.449	$&$	-3.467	$&$	$-$	$&$	$-$	$&		\\
Cr I	&$	4964.93	$&$	0.941	$&$	-2.527	$&$	3.58	$&$	$-$	$&		\\
Cr II	&$	4965.00	$&$	8.354	$&$	-5.841	$&$	$-$	$&$	$-$	$&		\\
Cr I	&$	6330.09	$&$	0.941	$&$	-2.920	$&$	3.56	$&$	$-$	$&		\\
Cr II	&$	6330.40	$&$	11.144	$&$	-1.000	$&$	$-$	$&$	$-$	$&		\\
\hline													
Mn I	&$	4055.54	$&$	2.143	$&$	-0.070	$&$	3.06	$&$	$-$	$&	blended	\\
Mn III	&$	4079.03	$&$	11.095	$&$	-9.881	$&$	2.88	$&$	$-$	$&	very blended	\\
Mn III	&$	4079.18	$&$	23.795	$&$	-0.432	$&$	$-$	$&$	$-$	$&		\\
Mn I	&$	4079.22	$&$	4.258	$&$	-0.161	$&$	$-$	$&$	$-$	$&		\\
Mn I	&$	4079.24	$&$	2.143	$&$	-0.530	$&$	$-$	$&$	$-$	$&		\\
Mn I	&$	4079.41	$&$	2.187	$&$	-0.420	$&$	$-$	$&$	$-$	$&		\\
Mn II	&$	4753.74	$&$	6.528	$&$	-4.469	$&$	3.30	$&$	$-$	$&		\\
Mn I	&$	4753.89	$&$	5.214	$&$	-0.965	$&$	$-$	$&$	$-$	$&		\\
Mn II	&$	4754.03	$&$	10.271	$&$	-3.379	$&$	$-$	$&$	$-$	$&		\\
Mn I	&$	4754.04	$&$	2.282	$&$	-0.086	$&$	$-$	$&$	$-$	$&		\\
Mn II	&$	4754.06	$&$	6.139	$&$	-3.081	$&$	$-$	$&$	$-$	$&		\\
Mn I	&$	4762.37	$&$	2.888	$&$	0.425	$&$	2.94	$&$	$-$	$&	blended	\\
Mn III	&$	4765.84	$&$	14.459	$&$	-4.448	$&$	3.22	$&$	$-$	$&		\\
Mn I	&$	4765.85	$&$	2.941	$&$	-0.080	$&$	$-$	$&$	$-$	$&		\\
Mn I	&$	4766.00	$&$	4.425	$&$	-1.730	$&$	$-$	$&$	$-$	$&		\\
Mn III	&$	4766.29	$&$	24.215	$&$	-1.865	$&$	3.16	$&$	$-$	$&		\\
Mn I	&$	4766.42	$&$	2.920	$&$	0.100	$&$	$-$	$&$	$-$	$&		\\
Mn III	&$	4766.46	$&$	21.564	$&$	-2.589	$&$	$-$	$&$	$-$	$&		\\
Mn II	&$	4766.52	$&$	10.661	$&$	-4.115	$&$	$-$	$&$	$-$	$&		\\
Mn I	&$	4766.66	$&$	5.199	$&$	-0.560	$&$	$-$	$&$	$-$	$&		\\
Mn II	&$	4783.29	$&$	10.283	$&$	-4.252	$&$	3.32	$&$	$-$	$&		\\
Mn I	&$	4783.43	$&$	2.298	$&$	0.042	$&$	$-$	$&$	$-$	$&		\\
Mn I	&$	4823.52	$&$	2.319	$&$	0.144	$&$	3.30	$&$	$-$	$&	blended	\\
Mn II	&$	4823.65	$&$	11.683	$&$	-2.969	$&$	$-$	$&$	$-$	$&		\\
Mn I	&$	6013.51	$&$	3.072	$&$	-0.251	$&$	3.00	$&$	$-$	$&		\\
\hline													
Fe I	&$	5762.84	$&$	4.301	$&$	-2.620	$&$	5.48	$&$	$-$	$&		\\
Fe I	&$	5762.98	$&$	4.191	$&$	-3.199	$&$	$-$	$&$	$-$	$&		\\
Fe I	&$	5762.99	$&$	4.209	$&$	-0.450	$&$	$-$	$&$	$-$	$&		\\
Fe I	&$	5763.00	$&$	4.191	$&$	-4.561	$&$	$-$	$&$	$-$	$&		\\
Fe I	&$	5862.23	$&$	3.640	$&$	-4.359	$&$	5.30	$&$	$-$	$&		\\
Fe I	&$	5862.36	$&$	4.549	$&$	-0.127	$&$	$-$	$&$	$-$	$&		\\
Fe I	&$	5862.47	$&$	5.334	$&$	-3.802	$&$	$-$	$&$	$-$	$&		\\
Fe I	&$	5914.11	$&$	4.608	$&$	-0.375	$&$	5.34	$&$	$-$	$&		\\
Fe I	&$	5914.20	$&$	4.608	$&$	-0.131	$&$	$-$	$&$	$-$	$&		\\
Fe I	&$	5956.41	$&$	5.352	$&$	-5.829	$&$	5.68	$&$	$-$	$&		\\
Fe I	&$	5956.46	$&$	4.283	$&$	-4.169	$&$	$-$	$&$	$-$	$&		\\
Fe I	&$	5956.51	$&$	5.352	$&$	-8.937	$&$	$-$	$&$	$-$	$&		\\
Fe I	&$	5956.69	$&$	0.859	$&$	-4.605	$&$	$-$	$&$	$-$	$&		\\
Fe I	&$	5956.94	$&$	4.580	$&$	-3.540	$&$	$-$	$&$	$-$	$&		\\
Fe I	&$	5976.69	$&$	5.330	$&$	-5.182	$&$	5.44	$&$	$-$	$&		\\
Fe I	&$	5976.78	$&$	3.943	$&$	-1.243	$&$	$-$	$&$	$-$	$&		\\
Fe I	&$	6002.77	$&$	5.314	$&$	-3.755	$&$	5.58	$&$	$-$	$&		\\
Fe I	&$	6002.79	$&$	5.386	$&$	-4.886	$&$	$-$	$&$	$-$	$&		\\
Fe I	&$	6003.01	$&$	3.881	$&$	-1.120	$&$	$-$	$&$	$-$	$&		\\
Fe I	&$	6003.02	$&$	5.506	$&$	-7.385	$&$	$-$	$&$	$-$	$&		\\
Fe I	&$	6003.11	$&$	5.506	$&$	-7.472	$&$	$-$	$&$	$-$	$&		\\
Fe I	&$	6003.11	$&$	5.506	$&$	-8.111	$&$	$-$	$&$	$-$	$&		\\
Fe I	&$	6003.17	$&$	5.506	$&$	-7.907	$&$	$-$	$&$	$-$	$&		\\
Fe I	&$	6055.76	$&$	5.352	$&$	-7.003	$&$	5.58	$&$	$-$	$&		\\
Fe I	&$	6055.78	$&$	5.357	$&$	-7.928	$&$	$-$	$&$	$-$	$&		\\
Fe I	&$	6055.89	$&$	5.352	$&$	-7.777	$&$	$-$	$&$	$-$	$&		\\
Fe I	&$	6055.94	$&$	5.070	$&$	-2.322	$&$	$-$	$&$	$-$	$&		\\
Fe I	&$	6055.94	$&$	5.352	$&$	-5.833	$&$	$-$	$&$	$-$	$&		\\
Fe I	&$	6055.95	$&$	5.357	$&$	-6.627	$&$	$-$	$&$	$-$	$&		\\
Fe I	&$	6055.95	$&$	5.357	$&$	-8.619	$&$	$-$	$&$	$-$	$&		\\
Fe I	&$	6056.01	$&$	4.733	$&$	-0.460	$&$	$-$	$&$	$-$	$&		\\
Fe I	&$	6056.09	$&$	5.357	$&$	-7.290	$&$	$-$	$&$	$-$	$&		\\
Fe I	&$	6056.09	$&$	5.357	$&$	-7.768	$&$	$-$	$&$	$-$	$&		\\
Fe I	&$	6056.25	$&$	5.357	$&$	-7.202	$&$	$-$	$&$	$-$	$&		\\
Fe I	&$	6065.48	$&$	2.608	$&$	-1.530	$&$	5.48	$&$	$-$	$&		\\
Fe I	&$	6065.49	$&$	4.956	$&$	-3.471	$&$	$-$	$&$	$-$	$&		\\
Fe I	&$	6082.46	$&$	5.486	$&$	-8.673	$&$	5.54	$&$	$-$	$&		\\
Fe I	&$	6082.54	$&$	5.386	$&$	-5.272	$&$	$-$	$&$	$-$	$&		\\
Fe I	&$	6082.58	$&$	5.607	$&$	-9.656	$&$	$-$	$&$	$-$	$&		\\
Fe I	&$	6082.58	$&$	5.607	$&$	-3.403	$&$	$-$	$&$	$-$	$&		\\
Fe I	&$	6082.61	$&$	4.220	$&$	-3.746	$&$	$-$	$&$	$-$	$&		\\
Fe I	&$	6082.71	$&$	2.223	$&$	-3.573	$&$	$-$	$&$	$-$	$&		\\
Fe I	&$	6082.85	$&$	5.341	$&$	-3.667	$&$	$-$	$&$	$-$	$&		\\
Fe I	&$	6082.89	$&$	5.486	$&$	-8.142	$&$	$-$	$&$	$-$	$&		\\
Fe I	&$	6082.89	$&$	5.486	$&$	-9.294	$&$	$-$	$&$	$-$	$&		\\
Fe I	&$	6136.25	$&$	5.273	$&$	-3.556	$&$	5.48	$&$	$-$	$&		\\
Fe I	&$	6136.62	$&$	2.453	$&$	-1.400	$&$	$-$	$&$	$-$	$&		\\
Fe I	&$	6136.99	$&$	2.198	$&$	-2.950	$&$	$-$	$&$	$-$	$&		\\
Fe I	&$	6137.28	$&$	4.580	$&$	-2.160	$&$	$-$	$&$	$-$	$&		\\
Fe I	&$	6137.47	$&$	4.301	$&$	-3.741	$&$	$-$	$&$	$-$	$&		\\
Fe I	&$	6137.50	$&$	3.332	$&$	-2.514	$&$	$-$	$&$	$-$	$&		\\
Fe I	&$	6137.69	$&$	2.588	$&$	-1.403	$&$	$-$	$&$	$-$	$&		\\
Fe I	&$	6151.62	$&$	2.176	$&$	-3.299	$&$	5.58	$&$	$-$	$&		\\
Fe I	&$	6151.69	$&$	5.012	$&$	-3.761	$&$	$-$	$&$	$-$	$&		\\
Fe I	&$	6173.01	$&$	0.990	$&$	-7.794	$&$	5.62	$&$	$-$	$&		\\
Fe I	&$	6173.03	$&$	3.640	$&$	-4.961	$&$	$-$	$&$	$-$	$&		\\
Fe I	&$	6173.15	$&$	4.991	$&$	-3.920	$&$	$-$	$&$	$-$	$&		\\
Fe I	&$	6173.29	$&$	5.372	$&$	-5.600	$&$	$-$	$&$	$-$	$&		\\
Fe I	&$	6173.33	$&$	2.223	$&$	-2.880	$&$	$-$	$&$	$-$	$&		\\
Fe I	&$	6173.34	$&$	5.334	$&$	-4.314	$&$	$-$	$&$	$-$	$&		\\
Fe I	&$	6173.49	$&$	5.348	$&$	-7.269	$&$	$-$	$&$	$-$	$&		\\
Fe I	&$	6173.64	$&$	4.446	$&$	-3.400	$&$	$-$	$&$	$-$	$&		\\
Fe I	&$	6180.20	$&$	2.727	$&$	-2.586	$&$	5.52	$&$	$-$	$&		\\
Fe I	&$	6180.29	$&$	5.314	$&$	-4.453	$&$	$-$	$&$	$-$	$&		\\
Fe I	&$	6191.56	$&$	2.433	$&$	-1.417	$&$	5.30	$&$	$-$	$&		\\
Fe I	&$	6191.57	$&$	4.256	$&$	-5.483	$&$	$-$	$&$	$-$	$&		\\
Fe I	&$	6191.77	$&$	4.143	$&$	-5.123	$&$	$-$	$&$	$-$	$&		\\
Fe I	&$	6200.27	$&$	4.320	$&$	-3.931	$&$	5.54	$&$	$-$	$&		\\
Fe I	&$	6200.31	$&$	2.608	$&$	-2.437	$&$	$-$	$&$	$-$	$&		\\
Fe I	&$	6200.51	$&$	5.357	$&$	-9.018	$&$	$-$	$&$	$-$	$&		\\
Fe I	&$	6213.27	$&$	5.386	$&$	-4.197	$&$	5.52	$&$	$-$	$&		\\
Fe I	&$	6213.43	$&$	2.223	$&$	-2.482	$&$	$-$	$&$	$-$	$&		\\
Fe I	&$	6218.94	$&$	5.446	$&$	-8.948	$&$	5.56	$&$	$-$	$&		\\
Fe I	&$	6219.14	$&$	5.010	$&$	-2.270	$&$	$-$	$&$	$-$	$&		\\
Fe I	&$	6219.28	$&$	2.198	$&$	-2.433	$&$	$-$	$&$	$-$	$&		\\
Fe I	&$	6219.31	$&$	5.458	$&$	-5.082	$&$	$-$	$&$	$-$	$&		\\
Fe I	&$	6219.53	$&$	3.417	$&$	-4.100	$&$	$-$	$&$	$-$	$&		\\
Fe I	&$	6219.58	$&$	5.345	$&$	-4.691	$&$	$-$	$&$	$-$	$&		\\
Fe I	&$	6230.48	$&$	5.410	$&$	-5.491	$&$	5.44	$&$	$-$	$&		\\
Fe I	&$	6230.72	$&$	2.559	$&$	-1.281	$&$	$-$	$&$	$-$	$&		\\
Fe I	&$	6252.38	$&$	4.320	$&$	-7.012	$&$	5.48	$&$	$-$	$&		\\
Fe I	&$	6252.56	$&$	2.404	$&$	-1.687	$&$	$-$	$&$	$-$	$&		\\
Fe I	&$	6252.78	$&$	5.648	$&$	-8.652	$&$	$-$	$&$	$-$	$&		\\
Fe I	&$	6254.26	$&$	2.279	$&$	-2.443	$&$	5.58	$&$	$-$	$&		\\
Fe I	&$	6301.50	$&$	3.654	$&$	-0.718	$&$	5.38	$&$	$-$	$&		\\
Fe I	&$	6301.77	$&$	5.458	$&$	-2.889	$&$	$-$	$&$	$-$	$&		\\
Fe I	&$	6317.76	$&$	5.314	$&$	-5.205	$&$	5.76	$&$	$-$	$&		\\
Fe I	&$	6318.02	$&$	2.453	$&$	-2.261	$&$	$-$	$&$	$-$	$&		\\
Fe I	&$	6318.04	$&$	5.683	$&$	-9.049	$&$	$-$	$&$	$-$	$&		\\
Fe I	&$	6318.05	$&$	5.683	$&$	-5.610	$&$	$-$	$&$	$-$	$&		\\
Fe I	&$	6318.05	$&$	5.683	$&$	-8.040	$&$	$-$	$&$	$-$	$&		\\
Fe I	&$	6318.35	$&$	5.314	$&$	-5.989	$&$	$-$	$&$	$-$	$&		\\
Fe I	&$	6322.69	$&$	2.588	$&$	-2.426	$&$	5.54	$&$	$-$	$&		\\
Fe I	&$	6322.74	$&$	5.491	$&$	-5.869	$&$	$-$	$&$	$-$	$&		\\
Fe I	&$	6335.33	$&$	2.198	$&$	-2.177	$&$	5.44	$&$	$-$	$&		\\
Fe I	&$	6344.02	$&$	4.415	$&$	-3.572	$&$	5.64	$&$	$-$	$&		\\
Fe I	&$	6344.07	$&$	5.524	$&$	-8.598	$&$	$-$	$&$	$-$	$&		\\
Fe I	&$	6344.08	$&$	5.524	$&$	-6.712	$&$	$-$	$&$	$-$	$&		\\
Fe I	&$	6344.15	$&$	2.433	$&$	-2.923	$&$	$-$	$&$	$-$	$&		\\
Fe I	&$	6344.28	$&$	5.486	$&$	-8.925	$&$	$-$	$&$	$-$	$&		\\
Fe I	&$	6344.28	$&$	4.473	$&$	-6.227	$&$	$-$	$&$	$-$	$&		\\
Fe I	&$	6344.37	$&$	5.486	$&$	-9.075	$&$	$-$	$&$	$-$	$&		\\
Fe I	&$	6344.37	$&$	5.486	$&$	-8.548	$&$	$-$	$&$	$-$	$&		\\
Fe I	&$	6358.43	$&$	4.593	$&$	-3.620	$&$	5.76	$&$	$-$	$&		\\
Fe I	&$	6358.51	$&$	5.388	$&$	-4.570	$&$	$-$	$&$	$-$	$&		\\
Fe I	&$	6358.63	$&$	5.345	$&$	-5.414	$&$	$-$	$&$	$-$	$&		\\
Fe I	&$	6358.63	$&$	4.143	$&$	-1.657	$&$	$-$	$&$	$-$	$&		\\
Fe I	&$	6358.65	$&$	4.371	$&$	-3.448	$&$	$-$	$&$	$-$	$&		\\
Fe I	&$	6358.70	$&$	0.859	$&$	-4.468	$&$	$-$	$&$	$-$	$&		\\
Fe I	&$	6399.53	$&$	5.669	$&$	-9.756	$&$	5.46	$&$	$-$	$&		\\
Fe I	&$	6399.61	$&$	5.669	$&$	-8.100	$&$	$-$	$&$	$-$	$&		\\
Fe I	&$	6399.61	$&$	5.669	$&$	-9.605	$&$	$-$	$&$	$-$	$&		\\
Fe I	&$	6399.65	$&$	3.984	$&$	-3.287	$&$	$-$	$&$	$-$	$&		\\
Fe I	&$	6399.69	$&$	5.502	$&$	-5.292	$&$	$-$	$&$	$-$	$&		\\
Fe I	&$	6399.77	$&$	5.669	$&$	-9.938	$&$	$-$	$&$	$-$	$&		\\
Fe I	&$	6399.79	$&$	5.357	$&$	-6.843	$&$	$-$	$&$	$-$	$&		\\
Fe I	&$	6399.85	$&$	5.502	$&$	-6.988	$&$	$-$	$&$	$-$	$&		\\
Fe I	&$	6400.00	$&$	3.602	$&$	-0.290	$&$	$-$	$&$	$-$	$&		\\
Fe I	&$	6400.06	$&$	5.314	$&$	-4.911	$&$	$-$	$&$	$-$	$&		\\
Fe I	&$	6400.13	$&$	5.669	$&$	-7.575	$&$	$-$	$&$	$-$	$&		\\
Fe I	&$	6400.32	$&$	0.915	$&$	-4.318	$&$	$-$	$&$	$-$	$&		\\
Fe I	&$	6400.34	$&$	5.064	$&$	-3.635	$&$	$-$	$&$	$-$	$&		\\
Fe I	&$	6400.37	$&$	5.502	$&$	-7.493	$&$	$-$	$&$	$-$	$&		\\
Fe I	&$	6400.37	$&$	5.502	$&$	-8.377	$&$	$-$	$&$	$-$	$&		\\
Fe I	&$	6411.65	$&$	3.654	$&$	-0.595	$&$	5.42	$&$	$-$	$&	\\
Fe I	&$	6411.99	$&$	5.446	$&$	-5.559	$&$	$-$	$&$	$-$	$&		\\
Fe I	&$	6421.21	$&$	5.334	$&$	-3.994	$&$	5.52	$&$	$-$	$&	\\
Fe I	&$	6421.35	$&$	2.279	$&$	-2.027	$&$	$-$	$&$	$-$	$&		\\
Fe I	&$	6421.57	$&$	5.334	$&$	-5.293	$&$	$-$	$&$	$-$	$&		\\
Fe I	&$	6494.98	$&$	2.404	$&$	-1.273	$&$	5.34	$&$	$-$	$&	\\
Fe I	&$	6545.85	$&$	4.580	$&$	-3.977	$&$	5.30	$&$	$-$	$&	\\
Fe I	&$	6545.99	$&$	5.314	$&$	-3.547	$&$	$-$	$&$	$-$	$&		\\
Fe I	&$	6546.20	$&$	5.357	$&$	-7.253	$&$	$-$	$&$	$-$	$&		\\
Fe I	&$	6546.24	$&$	2.758	$&$	-1.536	$&$	$-$	$&$	$-$	$&		\\
Fe I	&$	6546.47	$&$	5.841	$&$	-7.040	$&$	$-$	$&$	$-$	$&		\\
Fe I	&$	6546.50	$&$	4.473	$&$	-6.658	$&$	$-$	$&$	$-$	$&		\\
Fe I	&$	6592.61	$&$	4.956	$&$	-1.262	$&$	5.30	$&$	$-$	$&		\\
Fe I	&$	6592.91	$&$	2.727	$&$	-1.473	$&$	$-$	$&$	$-$	$&		\\
Fe I	&$	6593.87	$&$	2.433	$&$	-2.422	$&$	5.50	$&$	$-$	$&		\\
Fe I	&$	6608.03	$&$	2.279	$&$	-4.030	$&$	5.52	$&$	$-$	$&		\\
Fe I	&$	7494.72	$&$	1.557	$&$	-5.254	$&$	5.26	$&$	$-$	$&		\\
Fe I	&$	7494.84	$&$	5.458	$&$	-3.505	$&$	$-$	$&$	$-$	$&		\\
Fe I	&$	7494.90	$&$	4.985	$&$	-3.470	$&$	$-$	$&$	$-$	$&		\\
Fe I	&$	7495.00	$&$	3.695	$&$	-6.916	$&$	$-$	$&$	$-$	$&		\\
Fe I	&$	7495.01	$&$	5.720	$&$	-2.463	$&$	$-$	$&$	$-$	$&		\\
Fe I	&$	7495.07	$&$	4.220	$&$	0.053	$&$	$-$	$&$	$-$	$&		\\
Fe I	&$	7495.37	$&$	5.693	$&$	-3.673	$&$	$-$	$&$	$-$	$&		\\
Fe I	&$	7664.16	$&$	4.835	$&$	-1.176	$&$	5.42	$&$	$-$	$&		\\
Fe I	&$	7664.29	$&$	2.990	$&$	-1.683	$&$	$-$	$&$	$-$	$&		\\
Fe I	&$	7664.42	$&$	5.849	$&$	-2.832	$&$	$-$	$&$	$-$	$&		\\
Fe I	&$	7664.45	$&$	4.733	$&$	-3.527	$&$	$-$	$&$	$-$	$&		\\
Fe I	&$	7722.91	$&$	5.539	$&$	-5.064	$&$	5.78	$&$	$-$	$&		\\
Fe I	&$	7723.21	$&$	2.279	$&$	-3.617	$&$	$-$	$&$	$-$	$&		\\
Fe I	&$	7723.54	$&$	5.793	$&$	-4.026	$&$	$-$	$&$	$-$	$&		\\
Fe I	&$	7723.59	$&$	5.539	$&$	-5.366	$&$	$-$	$&$	$-$	$&		\\
Fe I	&$	7780.23	$&$	5.519	$&$	-3.671	$&$	5.24	$&$	$-$	$&		\\
Fe I	&$	7780.51	$&$	5.683	$&$	-3.647	$&$	$-$	$&$	$-$	$&		\\
Fe I	&$	7780.56	$&$	4.473	$&$	0.029	$&$	$-$	$&$	$-$	$&		\\
Fe I	&$	7780.59	$&$	5.693	$&$	-4.952	$&$	$-$	$&$	$-$	$&		\\
Fe I	&$	7780.75	$&$	5.930	$&$	-8.656	$&$	$-$	$&$	$-$	$&		\\
Fe I	&$	7831.86	$&$	5.386	$&$	-4.697	$&$	5.34	$&$	$-$	$&		\\
Fe I	&$	7832.20	$&$	4.435	$&$	0.111	$&$	$-$	$&$	$-$	$&		\\
Fe I	&$	8046.05	$&$	4.415	$&$	0.031	$&$	5.28	$&$	$-$	$&		\\
Fe I	&$	8387.77	$&$	2.176	$&$	-1.493	$&$	5.42	$&$	$-$	$&		\\
Fe I	&$	8387.96	$&$	5.879	$&$	-7.512	$&$	$-$	$&$	$-$	$&		\\
Fe I	&$	8388.07	$&$	5.642	$&$	-3.588	$&$	$-$	$&$	$-$	$&		\\
Fe I	&$	8515.11	$&$	3.018	$&$	-2.073	$&$	5.66	$&$	$-$	$&		\\
Fe I	&$	8621.60	$&$	2.949	$&$	-2.321	$&$	5.46	$&$	$-$	$&		\\
Fe I	&$	8621.89	$&$	5.967	$&$	-7.684	$&$	$-$	$&$	$-$	$&		\\
Fe I	&$	8674.31	$&$	5.720	$&$	-6.667	$&$	5.58	$&$	$-$	$&		\\
Fe I	&$	8674.36	$&$	6.010	$&$	-7.879	$&$	$-$	$&$	$-$	$&		\\
Fe I	&$	8674.54	$&$	5.996	$&$	-8.466	$&$	$-$	$&$	$-$	$&		\\
Fe I	&$	8674.58	$&$	5.621	$&$	-6.068	$&$	$-$	$&$	$-$	$&		\\
Fe I	&$	8674.58	$&$	5.693	$&$	-4.781	$&$	$-$	$&$	$-$	$&		\\
Fe I	&$	8674.75	$&$	2.831	$&$	-1.800	$&$	$-$	$&$	$-$	$&		\\
Fe I	&$	8675.19	$&$	5.913	$&$	-5.414	$&$	$-$	$&$	$-$	$&		\\
Fe I	&$	8763.97	$&$	4.652	$&$	-0.146	$&$	5.38	$&$	$-$	$&		\\
Fe I	&$	8975.11	$&$	5.979	$&$	-5.119	$&$	5.48	$&$	$-$	$&		\\
Fe I	&$	8975.16	$&$	4.988	$&$	-2.087	$&$	$-$	$&$	$-$	$&		\\
Fe I	&$	8975.27	$&$	3.686	$&$	-7.036	$&$	$-$	$&$	$-$	$&		\\
Fe I	&$	8975.31	$&$	4.143	$&$	-4.759	$&$	$-$	$&$	$-$	$&		\\
Fe I	&$	8975.40	$&$	2.990	$&$	-2.233	$&$	$-$	$&$	$-$	$&		\\
Fe I	&$	8975.42	$&$	5.883	$&$	-5.666	$&$	$-$	$&$	$-$	$&		\\
Fe I	&$	8975.69	$&$	5.334	$&$	-6.162	$&$	$-$	$&$	$-$	$&		\\
Fe I	&$	8975.80	$&$	5.352	$&$	-8.779	$&$	$-$	$&$	$-$	$&		\\
Fe I	&$	8999.56	$&$	2.831	$&$	-1.321	$&$	5.40	$&$	$-$	$&		\\
Fe I	&$	9088.03	$&$	5.502	$&$	-4.483	$&$	5.40	$&$	$-$	$&		\\
Fe I	&$	9088.32	$&$	2.845	$&$	-1.986	$&$	$-$	$&$	$-$	$&		\\
Fe I	&$	9088.35	$&$	5.930	$&$	-5.925	$&$	$-$	$&$	$-$	$&		\\
Fe I	&$	9088.46	$&$	6.010	$&$	-5.111	$&$	$-$	$&$	$-$	$&		\\
Fe I	&$	9088.50	$&$	5.928	$&$	-3.861	$&$	$-$	$&$	$-$	$&		\\
Fe I	&$	9089.40	$&$	2.949	$&$	-1.675	$&$	$-$	$&$	$-$	$&		\\
Fe I	&$	9089.50	$&$	5.947	$&$	-4.040	$&$	$-$	$&$	$-$	$&		\\
Fe I	&$	9145.78	$&$	6.252	$&$	-9.640	$&$	5.22	$&$	$-$	$&		\\
Fe I	&$	9145.80	$&$	5.693	$&$	-5.595	$&$	$-$	$&$	$-$	$&		\\
Fe I	&$	9146.13	$&$	2.588	$&$	-2.804	$&$	$-$	$&$	$-$	$&		\\
\hline													
Fe II	&$	6247.14	$&$	11.164	$&$	-4.093	$&$	5.70	$&$	0.14	$&	blended	\\
Fe II	&$	6247.26	$&$	11.237	$&$	-4.082	$&$	$-$	$&$	$-$	$&		\\
Fe II	&$	6247.35	$&$	6.209	$&$	-2.172	$&$	$-$	$&$	$-$	$&		\\
Fe II	&$	6247.37	$&$	10.909	$&$	-1.382	$&$	$-$	$&$	$-$	$&		\\
Fe II	&$	6247.44	$&$	12.276	$&$	-5.381	$&$	$-$	$&$	$-$	$&		\\
Fe II	&$	6247.56	$&$	3.892	$&$	-2.435	$&$	$-$	$&$	$-$	$&		\\
Fe II	&$	6247.57	$&$	5.956	$&$	-4.827	$&$	$-$	$&$	$-$	$&		\\
Fe II	&$	6247.66	$&$	11.591	$&$	-8.222	$&$	$-$	$&$	$-$	$&		\\
Fe II	&$	6432.47	$&$	10.448	$&$	-9.829	$&$	5.72	$&$	0.10	$&	blended	\\
Fe II	&$	6432.68	$&$	10.930	$&$	-1.236	$&$	$-$	$&$	$-$	$&		\\
Fe II	&$	6432.68	$&$	2.891	$&$	-3.687	$&$	$-$	$&$	$-$	$&		\\
Fe II	&$	6456.38	$&$	3.903	$&$	-2.185	$&$	5.64	$&$	0.12	$&	blended	\\
Fe II	&$	6456.47	$&$	11.547	$&$	-4.091	$&$	$-$	$&$	$-$	$&		\\
Fe II	&$	6516.08	$&$	2.891	$&$	-3.432	$&$	5.56	$&$	0.08	$&	blended	\\
\hline													
Co I	&$	4233.71	$&$	3.930	$&$	-3.106	$&$	3.02	$&$	$-$	$&		\\
Co I	&$	4233.98	$&$	0.000	$&$	-3.469	$&$	$-$	$&$	$-$	$&		\\
Co I	&$	4813.45	$&$	2.871	$&$	-2.121	$&$	3.28	$&$	$-$	$&		\\
Co I	&$	4813.47	$&$	3.216	$&$	0.050	$&$	$-$	$&$	$-$	$&		\\
Co 4	&$	4813.71	$&$	17.943	$&$	-6.686	$&$	$-$	$&$	$-$	$&		\\
Co II	&$	4840.10	$&$	11.326	$&$	-2.988	$&$	3.16	$&$	$-$	$&		\\
Co I	&$	4840.25	$&$	3.170	$&$	0.144	$&$	$-$	$&$	$-$	$&		\\
Co II	&$	4840.37	$&$	3.459	$&$	-5.518	$&$	$-$	$&$	$-$	$&		\\
Co I	&$	6450.08	$&$	2.137	$&$	-2.132	$&$	2.96	$&$	$-$	$&		\\
Co I	&$	6450.25	$&$	1.710	$&$	-1.698	$&$	$-$	$&$	$-$	$&		\\
Co I	&$	6872.39	$&$	2.008	$&$	-1.589	$&$	2.80	$&$	$-$	$&		\\
Co I	&$	7052.87	$&$	1.956	$&$	-1.264	$&$	2.58	$&$	$-$	$&		\\
Co I	&$	7084.98	$&$	1.883	$&$	-1.018	$&$	2.62	$&$	$-$	$&		\\
\hline													
Ni I	&$	5711.88	$&$	1.935	$&$	-2.270	$&$	4.18	$&$	$-$	$&	blended	\\
Ni I	&$	5715.07	$&$	4.088	$&$	-0.352	$&$	4.22	$&$	$-$	$&	blended	\\
Ni II	&$	5748.25	$&$	14.896	$&$	-0.858	$&$	4.10	$&$	$-$	$&		\\
Ni I	&$	5748.35	$&$	1.676	$&$	-3.260	$&$	$-$	$&$	$-$	$&		\\
Ni I	&$	5754.51	$&$	3.941	$&$	-3.611	$&$	4.38	$&$	$-$	$&		\\
Ni II	&$	5754.51	$&$	14.733	$&$	-2.531	$&$	$-$	$&$	$-$	$&		\\
Ni I	&$	5754.65	$&$	1.935	$&$	-2.330	$&$	$-$	$&$	$-$	$&		\\
Ni III	&$	5754.87	$&$	22.952	$&$	-2.744	$&$	$-$	$&$	$-$	$&		\\
Ni II	&$	5846.77	$&$	12.208	$&$	-3.583	$&$	4.14	$&$	$-$	$&		\\
Ni I	&$	5846.99	$&$	1.676	$&$	-3.210	$&$	$-$	$&$	$-$	$&		\\
Ni I	&$	5892.75	$&$	4.154	$&$	-1.494	$&$	4.32	$&$	$-$	$&	blended	\\
Ni I	&$	5892.87	$&$	1.986	$&$	-2.350	$&$	$-$	$&$	$-$	$&		\\
Ni II	&$	5892.94	$&$	14.667	$&$	-2.139	$&$	$-$	$&$	$-$	$&		\\
Ni II	&$	5893.25	$&$	15.008	$&$	-1.827	$&$	$-$	$&$	$-$	$&		\\
Ni I	&$	6007.31	$&$	1.676	$&$	-3.330	$&$	4.14	$&$	$-$	$&		\\
Ni II	&$	6007.35	$&$	13.080	$&$	-1.934	$&$	$-$	$&$	$-$	$&		\\
Ni I	&$	6108.11	$&$	1.676	$&$	-2.450	$&$	4.06	$&$	$-$	$&		\\
Ni I	&$	6116.17	$&$	4.089	$&$	-0.677	$&$	4.26	$&$	$-$	$&		\\
Ni I	&$	6116.17	$&$	4.266	$&$	-0.822	$&$	$-$	$&$	$-$	$&		\\
Ni II	&$	6116.20	$&$	15.026	$&$	-2.610	$&$	$-$	$&$	$-$	$&		\\
Ni I	&$	6128.96	$&$	1.676	$&$	-3.330	$&$	4.18	$&$	$-$	$&		\\
Ni II	&$	6176.62	$&$	15.008	$&$	-2.753	$&$	4.16	$&$	$-$	$&		\\
Ni I	&$	6176.81	$&$	4.088	$&$	-0.260	$&$	$-$	$&$	$-$	$&		\\
Ni I	&$	6177.24	$&$	1.826	$&$	-3.500	$&$	$-$	$&$	$-$	$&		\\
Ni II	&$	6177.28	$&$	11.969	$&$	-3.167	$&$	$-$	$&$	$-$	$&		\\
Ni I	&$	6177.54	$&$	4.236	$&$	-2.141	$&$	$-$	$&$	$-$	$&		\\
Ni II	&$	6314.46	$&$	12.904	$&$	0.480	$&$	3.62	$&$	$-$	$&		\\
Ni II	&$	6314.48	$&$	14.903	$&$	-1.870	$&$	$-$	$&$	$-$	$&		\\
Ni I	&$	6314.65	$&$	1.935	$&$	-1.770	$&$	$-$	$&$	$-$	$&		\\
Ni I	&$	6314.66	$&$	4.154	$&$	-0.921	$&$	$-$	$&$	$-$	$&		\\
Ni II	&$	6314.73	$&$	14.906	$&$	-2.400	$&$	$-$	$&$	$-$	$&		\\
Ni II	&$	6327.37	$&$	12.457	$&$	-3.303	$&$	4.14	$&$	$-$	$&		\\
Ni II	&$	6327.54	$&$	14.837	$&$	-3.471	$&$	$-$	$&$	$-$	$&		\\
Ni I	&$	6327.59	$&$	1.676	$&$	-3.150	$&$	$-$	$&$	$-$	$&		\\
Ni II	&$	6327.71	$&$	14.685	$&$	-1.895	$&$	$-$	$&$	$-$	$&		\\
Ni II	&$	6327.79	$&$	15.022	$&$	-1.043	$&$	$-$	$&$	$-$	$&		\\
Ni II	&$	6482.67	$&$	14.912	$&$	-1.094	$&$	3.94	$&$	$-$	$&		\\
Ni II	&$	6482.67	$&$	14.912	$&$	-1.010	$&$	$-$	$&$	$-$	$&		\\
Ni I	&$	6482.80	$&$	1.935	$&$	-2.630	$&$	$-$	$&$	$-$	$&		\\
Ni II	&$	6482.81	$&$	14.744	$&$	-1.692	$&$	$-$	$&$	$-$	$&		\\
Ni II	&$	6482.82	$&$	14.744	$&$	-1.185	$&$	$-$	$&$	$-$	$&		\\
Ni II	&$	6482.91	$&$	14.912	$&$	-2.596	$&$	$-$	$&$	$-$	$&		\\
Ni II	&$	6482.94	$&$	14.912	$&$	-2.779	$&$	$-$	$&$	$-$	$&		\\
Ni II	&$	6483.02	$&$	14.912	$&$	-2.258	$&$	$-$	$&$	$-$	$&		\\
Ni II	&$	6483.03	$&$	14.912	$&$	-2.782	$&$	$-$	$&$	$-$	$&		\\
Ni II	&$	6586.26	$&$	14.739	$&$	-2.772	$&$	4.10	$&$	$-$	$&		\\
Ni I	&$	6586.31	$&$	1.951	$&$	-2.810	$&$	$-$	$&$	$-$	$&		\\
Ni II	&$	6643.40	$&$	14.731	$&$	-3.243	$&$	4.26	$&$	$-$	$&		\\
Ni I	&$	6643.63	$&$	1.676	$&$	-2.300	$&$	$-$	$&$	$-$	$&		\\
Ni II	&$	6767.54	$&$	15.224	$&$	-2.851	$&$	4.20	$&$	$-$	$&		\\
Ni I	&$	6767.77	$&$	1.826	$&$	-2.170	$&$	$-$	$&$	$-$	$&		\\
Ni I	&$	6914.56	$&$	1.951	$&$	-2.270	$&$	4.12	$&$	$-$	$&		\\
Ni I	&$	7197.01	$&$	1.935	$&$	-2.680	$&$	4.28	$&$	$-$	$&	blended	\\
Ni II	&$	7197.25	$&$	14.334	$&$	-2.630	$&$	$-$	$&$	$-$	$&		\\
Ni I	&$	7197.39	$&$	5.004	$&$	-3.128	$&$	$-$	$&$	$-$	$&		\\
Ni II	&$	7393.23	$&$	15.059	$&$	-1.426	$&$	4.70	$&$	$-$	$&	blended	\\
Ni I	&$	7393.60	$&$	3.606	$&$	-0.825	$&$	$-$	$&$	$-$	$&		\\
Ni II	&$	7408.96	$&$	14.486	$&$	-3.296	$&$	4.04	$&$	$-$	$&	blended	\\
Ni II	&$	7409.01	$&$	14.473	$&$	-2.127	$&$	$-$	$&$	$-$	$&		\\
Ni I	&$	7409.04	$&$	5.497	$&$	-2.476	$&$	$-$	$&$	$-$	$&		\\
Ni I	&$	7409.25	$&$	5.514	$&$	-1.790	$&$	$-$	$&$	$-$	$&		\\
Ni I	&$	7409.35	$&$	3.796	$&$	-0.237	$&$	$-$	$&$	$-$	$&		\\
Ni I	&$	7414.50	$&$	1.986	$&$	-2.570	$&$	4.32	$&$	$-$	$&	blended	\\
\hline													
Cu I	&$	5782.13	$&$	1.642	$&$	-1.720	$&$	1.44	$&$	0.26	$&		\\
\hline													
Zn I	&$	4680.13	$&$	4.006	$&$	-0.815	$&$	2.82	$&$	0.26	$&	blended	\\
Zn I	&$	4722.15	$&$	4.030	$&$	-0.338	$&$	2.60	$&$	0.26	$&	blended	\\
Zn I	&$	4810.53	$&$	4.078	$&$	-0.137	$&$	2.64	$&$	0.20	$&		\\
\hline													
Ga I	&$	4172.04	$&$	0.102	$&$	-0.270	$&$	0.82	$&$	0.44	$&	very blended	\\
\hline													
Sr II	&$	4077.71	$&$	0.000	$&$	0.167	$&$	1.44	$&$	0.34	$&	very blended	\\
Sr I	&$	4607.33	$&$	0.000	$&$	-0.570	$&$	1.70	$&$	0.16	$&		\\
\hline													
Y II	&$	4682.32	$&$	0.409	$&$	-1.510	$&$	0.70	$&$	$-$	$&	blended	\\
Y II	&$	4823.30	$&$	0.992	$&$	-1.110	$&$	0.56	$&$	$-$	$&	blended	\\
Y II	&$	4854.86	$&$	0.992	$&$	-0.380	$&$	0.60	$&$	$-$	$&		\\
Y II	&$	4883.68	$&$	1.084	$&$	0.070	$&$	0.50	$&$	$-$	$&		\\
Y I	&$	4900.08	$&$	1.398	$&$	-0.360	$&$	0.52	$&$	$-$	$&	blended	\\
Y II	&$	4900.12	$&$	1.033	$&$	-0.090	$&$	$-$	$&$	$-$	$&		\\
Y I	&$	4981.97	$&$	1.983	$&$	-1.980	$&$	0.44	$&$	$-$	$&		\\
Y II	&$	4982.13	$&$	1.033	$&$	-1.290	$&$	$-$	$&$	$-$	$&		\\
\hline													
Zr II	&$	3998.95	$&$	0.559	$&$	-0.520	$&$	1.08	$&$	$-$	$&	blended	\\
Zr II	&$	4029.68	$&$	0.713	$&$	-0.780	$&$	0.98	$&$	$-$	$&	blended	\\
Zr II	&$	4258.04	$&$	0.559	$&$	-1.200	$&$	0.74	$&$	$-$	$&	blended	\\
Zr II	&$	4317.30	$&$	0.713	$&$	-1.450	$&$	0.92	$&$	$-$	$&		\\
Zr II	&$	4613.95	$&$	0.972	$&$	-1.540	$&$	0.98	$&$	$-$	$&		\\
\hline													
Ba II	&$	4554.03	$&$	0.000	$&$	0.170	$&$	-0.34	$&$	$-$	$&	blended	\\
Ba II	&$	4934.08	$&$	0.000	$&$	-0.150	$&$	-0.32	$&$	$-$	$&	blended	\\
Ba II	&$	5853.67	$&$	0.604	$&$	-1.000	$&$	-0.48	$&$	$-$	$&		\\
Ba II	&$	6141.71	$&$	0.704	$&$	-0.076	$&$	-0.44	$&$	$-$	$&		\\
Ba II	&$	6496.90	$&$	0.604	$&$	-0.377	$&$	-0.14	$&$	$-$	$&		\\
\hline													
La II	&$	3988.51	$&$	0.403	$&$	0.170	$&$	-1.34	$&$	$-$	$&		\\
La II	&$	3995.75	$&$	0.173	$&$	-0.100	$&$	-1.00	$&$	$-$	$&	blended	\\
La II	&$	4031.69	$&$	0.321	$&$	-0.090	$&$	-0.96	$&$	$-$	$&	very blended	\\
La II	&$	4086.71	$&$	0.000	$&$	-0.070	$&$	-1.24	$&$	$-$	$&	very blended	\\
La II	&$	4123.22	$&$	0.321	$&$	0.110	$&$	-0.54	$&$	$-$	$&	very blended	\\
La II	&$	4333.75	$&$	0.173	$&$	-0.060	$&$	-0.80	$&$	$-$	$&	very blended	\\
La II	&$	4662.50	$&$	0.000	$&$	-1.240	$&$	-1.12	$&$	$-$	$&		\\
La II	&$	4740.28	$&$	0.126	$&$	-1.050	$&$	-1.06	$&$	$-$	$&	blended	\\
La II	&$	4920.98	$&$	0.126	$&$	-0.580	$&$	-1.46	$&$	$-$	$&	blended	\\
La II	&$	4921.78	$&$	0.244	$&$	-0.450	$&$	-1.26	$&$	$-$	$&	blended	\\
\hline													
Ce II	&$	4053.50	$&$	0.000	$&$	-0.710	$&$	-0.62	$&$	$-$	$&	very blended	\\
Ce II	&$	4053.53	$&$	0.621	$&$	-2.740	$&$	$-$	$&$	$-$	$&		\\
Ce II	&$	4053.57	$&$	3.469	$&$	-2.910	$&$	$-$	$&$	$-$	$&		\\
Ce II	&$	4053.59	$&$	1.048	$&$	-1.490	$&$	$-$	$&$	$-$	$&		\\
Ce II	&$	4418.78	$&$	0.864	$&$	0.280	$&$	-0.20	$&$	$-$	$&		\\
Ce II	&$	4483.89	$&$	0.864	$&$	0.150	$&$	-0.22	$&$	$-$	$&	blended	\\
Ce III	&$	4484.06	$&$	12.490	$&$	-1.860	$&$	$-$	$&$	$-$	$&		\\
Ce II	&$	4486.88	$&$	1.339	$&$	-1.200	$&$	-0.56	$&$	$-$	$&		\\
Ce II	&$	4486.91	$&$	0.295	$&$	-0.260	$&$	$-$	$&$	$-$	$&		\\
Ce II	&$	4487.00	$&$	3.562	$&$	-1.100	$&$	$-$	$&$	$-$	$&		\\
Ce II	&$	4487.12	$&$	0.122	$&$	-3.350	$&$	$-$	$&$	$-$	$&		\\
Ce II	&$	4487.17	$&$	0.232	$&$	-2.540	$&$	$-$	$&$	$-$	$&		\\
Ce II	&$	4539.42	$&$	1.930	$&$	-0.810	$&$	-1.18	$&$	$-$	$&	blended	\\
Ce II	&$	4539.50	$&$	1.042	$&$	-3.320	$&$	$-$	$&$	$-$	$&		\\
Ce II	&$	4539.58	$&$	0.900	$&$	-1.060	$&$	$-$	$&$	$-$	$&		\\
Ce II	&$	4539.61	$&$	1.412	$&$	-0.910	$&$	$-$	$&$	$-$	$&		\\
Ce II	&$	4539.75	$&$	0.328	$&$	-0.020	$&$	$-$	$&$	$-$	$&		\\
Ce II	&$	4539.85	$&$	1.645	$&$	-2.050	$&$	$-$	$&$	$-$	$&		\\
Ce II	&$	4562.28	$&$	1.327	$&$	-2.120	$&$	-0.92	$&$	$-$	$&		\\
Ce II	&$	4562.36	$&$	0.478	$&$	0.230	$&$	$-$	$&$	$-$	$&		\\
Ce II	&$	4572.28	$&$	0.684	$&$	0.290	$&$	-0.56	$&$	$-$	$&	blended	\\
Ce II	&$	4628.16	$&$	0.516	$&$	0.200	$&$	-0.68	$&$	$-$	$&		\\
Ce II	&$	4628.19	$&$	1.194	$&$	-3.280	$&$	$-$	$&$	$-$	$&		\\
Ce II	&$	4628.24	$&$	1.366	$&$	-0.430	$&$	$-$	$&$	$-$	$&		\\
\hline													
Pr II	&$	4408.82	$&$	0.000	$&$	-0.278	$&$	<-1.30	$&$	$-$	$&	blended	\\
Pr II	&$	4408.82	$&$	0.000	$&$	-0.278	$&$	<-1.10	$&$	$-$	$&	blended	\\
\hline													
Nd II	&$	4004.00	$&$	0.064	$&$	-0.570	$&$	-0.58	$&$	$-$	$&	blended	\\
Nd II	&$	4021.33	$&$	0.321	$&$	-0.100	$&$	-0.74	$&$	$-$	$&		\\
Nd II	&$	4351.28	$&$	0.182	$&$	-0.610	$&$	-0.40	$&$	$-$	$&	very blended	\\
Nd II	&$	4446.38	$&$	0.205	$&$	-0.350	$&$	-0.84	$&$	$-$	$&		\\
Nd II	&$	4506.58	$&$	0.064	$&$	-1.040	$&$	-0.88	$&$	$-$	$&	blended	\\
Nd II	&$	4811.34	$&$	0.064	$&$	-1.015	$&$	-0.38	$&$	$-$	$&		\\
Nd II	&$	4825.48	$&$	0.182	$&$	-0.420	$&$	-0.62	$&$	$-$	$&		\\
Nd II	&$	4859.03	$&$	0.321	$&$	-0.440	$&$	-0.62	$&$	$-$	$&		\\
Nd II	&$	4959.12	$&$	0.064	$&$	-0.800	$&$	-0.44	$&$	$-$	$&		\\
Nd II	&$	6900.44	$&$	0.000	$&$	-1.542	$&$	-0.60	$&$	$-$	$&		\\
\hline													
Sm II	&$	4318.93	$&$	0.277	$&$	-0.250	$&$	-1.02	$&$	0.50	$&	blended	\\
Sm II	&$	4434.32	$&$	0.378	$&$	-0.070	$&$	-0.66	$&$	0.32	$&	blended	\\
Sm II	&$	4467.34	$&$	0.659	$&$	0.150	$&$	-1.04	$&$	0.26	$&	blended	\\
\hline													
Eu II	&$	3907.11	$&$	0.207	$&$	0.170	$&$	-1.72	$&$	0.38	$&	very blended	\\
Eu II	&$	4129.72	$&$	0.000	$&$	0.220	$&$	-1.44	$&$	0.38	$&	very blended	\\
Eu II	&$	4205.04	$&$	0.000	$&$	0.210	$&$	-1.66	$&$	0.66	$&	very blended	\\
Eu II	&$	4435.58	$&$	0.207	$&$	-0.110	$&$	-1.22	$&$	0.40	$&	very blended	\\
Eu II	&$	6645.06	$&$	1.380	$&$	0.120	$&$	<-1.22	$&$	$-$	$&		\\
\hline													
Gd II	&$	4251.73	$&$	0.382	$&$	-0.220	$&$	<-0.50	$&$	$-$	$&		\\
\hline													
Tb II	&$	4752.53	$&$	0.000	$&$	-0.550	$&$	<-1.38	$&$	$-$	$&	blended	\\
\hline													
Dy II	&$	3944.68	$&$	0.000	$&$	0.110	$&$	-0.86	$&$	0.50	$&	very blended	\\
Dy II	&$	3983.65	$&$	0.538	$&$	-0.310	$&$	<-0.38	$&$	$-$	$&		very blended \\
Dy II	&$	4103.31	$&$	0.103	$&$	-0.380	$&$	-1.00	$&$	0.56	$&	very blended	\\
Dy II	&$	4449.70	$&$	0.000	$&$	-1.030	$&$	<-0.66	$&$	$-$	$&		very blended \\
\hline													
Er II	&$	3896.23	$&$	0.055	$&$	-0.241	$&$	<-0.60	$&$	$-$	$&		very blended\\
\hline													
Pb I	&$	4057.81	$&$	1.320	$&$	-0.170	$&$	<0.92	$&$	$-$	$&		very blended\\
\hline													
Th II	&$	4250.34	$&$	0.557	$&$	0.000	$&$	<-1.20	$&$	$-$	$&		blended \\
\hline 
\label{table:linelist}
\end{longtable}
\end{longtab}

\begin{table}
\caption{Element abundances of ET0097. [Fe~I/H] is adopted as the reference iron value [Fe/H]. For other elements the adopted value is an average of all lines available, in some cases more than one ionization stage, see Table \ref{table:linelist} for details. No corrections have been made for NLTE effects. The solar values are adopted from \citet{GrevesseSauval1998}.}      
\label{table:abundances}      
\centering                          
\begin{tabular}{l r r r r r r}        
\hline        \hline
X &  $\log \epsilon_\odot$ &   $N_\textsl{X}$ & $\log \epsilon$  &  [X/Fe] & $\delta_\textsl{noise}$ &  $\delta_\textsl{total}$ \\    
\hline     
Fe~I		&$	7.50	$&$	49	$&$	5.47	$&$	-2.03\tablefootmark{a}	$&$	0.02	$&$	0.10	$\\
Fe~II	&$	7.50	$&$	4	$&$	5.64	$&$	-1.86\tablefootmark{a}	$&$	0.10	$&$	0.14	$\\
Li	&$	1.10	$&$	1	$&$	<0.17	$&$	<1.10	$&$	$-$	$&$	$-$	$\\
C	&$	8.52	$&$	$-$	$&$	7.00	$&$	0.51	$&$	0.09	$&$	0.10	$\\
N	&$	7.92	$&$	$-$	$&$	7.07	$&$	1.18	$&$	0.20	$&$	0.20	$\\
O	&$	8.83	$&$	2	$&$	7.33	$&$	0.53	$&$	0.18	$&$	0.19	$\\
Na	&$	6.33	$&$	3	$&$	3.81	$&$	-0.49	$&$	0.24	$&$	0.24	$\\
Mg	&$	7.58	$&$	5	$&$	5.97	$&$	0.42	$&$	0.04	$&$	0.06	$\\
Al	&$	6.47	$&$	1	$&$	3.62	$&$	-0.82	$&$	0.80	$&$	0.80	$\\
Si	&$	7.55	$&$	7	$&$	5.77	$&$	0.25	$&$	0.09	$&$	0.11	$\\
S	&$	7.33	$&$	2	$&$	5.44	$&$	0.14	$&$	0.29	$&$	0.29	$\\
K	&$	5.12	$&$	2	$&$	3.45	$&$	0.36	$&$	0.07	$&$	0.08	$\\
Ca	&$	6.36	$&$	20	$&$	4.49	$&$	0.16	$&$	0.02	$&$	0.04	$\\
Sc	&$	3.17	$&$	9	$&$	1.10	$&$	-0.04	$&$	0.04	$&$	0.06	$\\
Ti	&$	5.02	$&$	35	$&$	3.12	$&$	0.13	$&$	0.02	$&$	0.07	$\\
V	&$	4.00	$&$	10	$&$	2.01	$&$	0.04	$&$	0.04	$&$	0.06	$\\
Cr	&$	5.67	$&$	16	$&$	3.53	$&$	-0.11	$&$	0.07	$&$	0.08	$\\
Mn	&$	5.39	$&$	9	$&$	3.13	$&$	-0.23	$&$	0.06	$&$	0.07	$\\
Co	&$	4.92	$&$	7	$&$	2.92	$&$	0.03	$&$	0.10	$&$	0.11	$\\
Ni	&$	6.25	$&$	22	$&$	4.18	$&$	-0.04	$&$	0.04	$&$	0.07	$\\
Cu	&$	4.21	$&$	1	$&$	1.44	$&$	-0.74	$&$	0.26	$&$	0.26	$\\
Zn	&$	4.60	$&$	3	$&$	2.68	$&$	0.11	$&$	0.23	$&$	0.23	$\\
Ga	&$	2.88	$&$	1	$&$	0.82	$&$	-0.03	$&$	0.44	$&$	0.44	$\\
Sr	&$	2.97	$&$	2	$&$	1.65	$&$	0.71	$&$	0.20	$&$	0.20	$\\
Y	&$	2.24	$&$	6	$&$	0.55	$&$	0.34	$&$	0.04	$&$	0.06	$\\
Zr	&$	2.60	$&$	5	$&$	0.94	$&$	0.37	$&$	0.06	$&$	0.07	$\\
Ba	&$	2.13	$&$	5	$&$	-0.34	$&$	-0.44	$&$	0.06	$&$	0.08	$\\
La	&$	1.17	$&$	10	$&$	-1.08	$&$	-0.22	$&$	0.09	$&$	0.09	$\\
Ce	&$	1.58	$&$	8	$&$	-0.62	$&$	-0.17	$&$	0.12	$&$	0.13	$\\
Pr	&$	0.71	$&$	1	$&$	<-1.10	$&$	<0.22	$&$	$-$	$&$	$-$	$\\
Nd	&$	1.50	$&$	10	$&$	-0.61	$&$	-0.08	$&$	0.05	$&$	0.07	$\\
Sm	&$	1.01	$&$	3	$&$	-0.91	$&$	0.11	$&$	0.32	$&$	0.32	$\\
Eu	&$	0.51	$&$	4	$&$	-1.49	$&$	0.03	$&$	0.42	$&$	0.42	$\\
Gd	&$	1.12	$&$	1	$&$	<-0.50	$&$	<0.41	$&$	$-$	$&$	$-$	$\\
Tb	&$	0.35	$&$	1	$&$	<-1.38	$&$	<0.30	$&$	$-$	$&$	$-$	$\\
Dy	&$	1.14	$&$	2	$&$	-0.92	$&$	-0.03	$&$	0.53	$&$	0.53	$\\
Er	&$	0.93	$&$	1	$&$	<-0.60	$&$	<0.50	$&$	$-$	$&$	$-$	$\\
Pb	&$	1.95	$&$	1	$&$	<0.92	$&$	<1.00	$&$	$-$	$&$	$-$	$\\
Th	&$	0.09	$&$	1	$&$	<-1.20	$&$	<0.74	$&$	$-$	$&$	$-$	$\\
\hline 
\multicolumn{2}{l}{$^a$[Fe/H]} 
\end{tabular}
\end{table}

%
\section{Abundance measurements}

All analysis was carried out using the spectral synthesis code TURBOSPEC developed by Bernand Plez \citep{AlvarezPlez1998,Plez2012}. Because of the evolutionary stage of ET0097, its relatively low temperature and its high C and N abundances, most of the observed wavelength range is covered with CN and/or CH molecular lines. Synthetic spectra are therefore necessary to include the effects of these lines in the abundance evaluation and to distinguish between lines with different degrees of blending. The line list used was chosen to minimize the use of heavily blended lines, and when possible only lines with minimal or no blending were used (see Table \ref{table:linelist}).

The stellar atmosphere models are adopted from MARCS \citep{Gustafsson2008} for stars with standard composition, 1D and assuming local thermodynamic equilibrium (LTE), interpolated to match the exact stellar parameters for ET0097. Parameters for the atomic lines are adopted from the DREAM data base \citep{Biemont1999}, extracted via VALD (\citealt{Kupka1999} and references therein). The molecular parameters for C$_2$ come from M. Querci (private communication), and are described in \citet{Querci1972}. For CH molecular lines, we use the data from \citet{Plez2007}. The molecular parameters for CN are from T. Masseron (private communication), derived with similar methods and lab data as \citet{Brooke2014} and \citet{Sneden2014}.

The results of the abundance analysis are listed in Table~\ref{table:abundances}. To maintain consistency with Hill et al. (in prep.), the solar values used are from \citet{GrevesseSauval1998}, and where data from the literature is included, it is shifted to match this scale.

\subsection{Iron}
Special care was taken to ensure that the measured Fe lines were not severely blended with CN, C$_2$, or CH molecular lines. Only Fe~I lines that changed by less than 0.1 dex, when both the assumed carbon and nitrogen abundances were either increased or decreased by 0.2~dex, were used. Increasing both elements by this value had a strong effect on the CN line strength, so the real effect from the uncertainty on C and N abundance measurements should be much less than 0.1~dex on the selected Fe~I lines and, in all cases, it is less than the statistical errors on the individual lines. In total, 49 Fe~I lines could be used in the wavelength range 5600-9200~\text{\AA}, see Table \ref{table:linelist}. All Fe~I lines in the bluer part of the spectrum were too blended to comply with the rather strict criteria for unblended lines.

For Fe~II, no lines were found that fulfill the criteria, so the four least blended Fe~II lines were used for the measurement.

   \begin{figure} [!ht]
   \centering
   \includegraphics[width=\hsize]{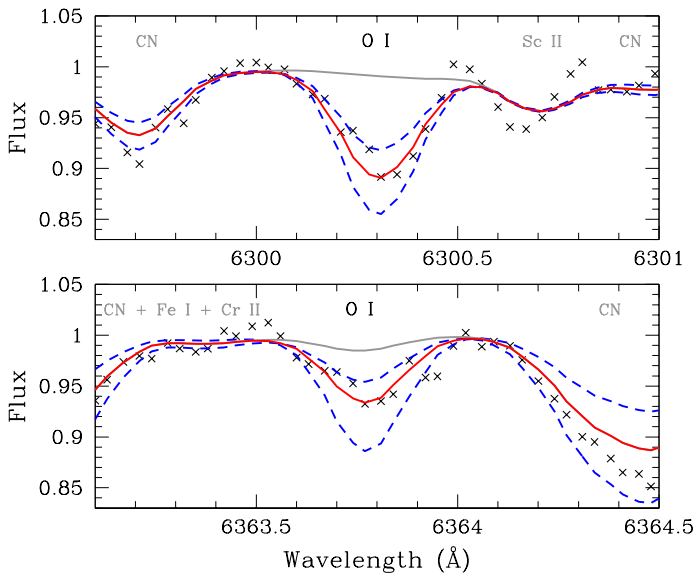}
      \caption{Measured oxygen lines, $\log \epsilon (\text{O})_{6300\text{\AA}}~=~7.30\pm0.14$ and $\log \epsilon (\text{O})_{6364\text{\AA}}~=~7.46\pm0.30$. Solid red lines show best fits, and dashed blue lines show upper and lower limits of error bars. Cases where [O~I] lines have been removed from the linelist are shown with solid gray lines. The strength of CN molecular lines is very sensitive to oxygen abundance. Adopting a lower O abundance will make CN lines in the synthetic spectrum stronger, assuming fixed values of C and N.
      }
         \label{O}
   \end{figure}

      \begin{figure}
   \centering
   \includegraphics[width=\hsize-0.4cm]{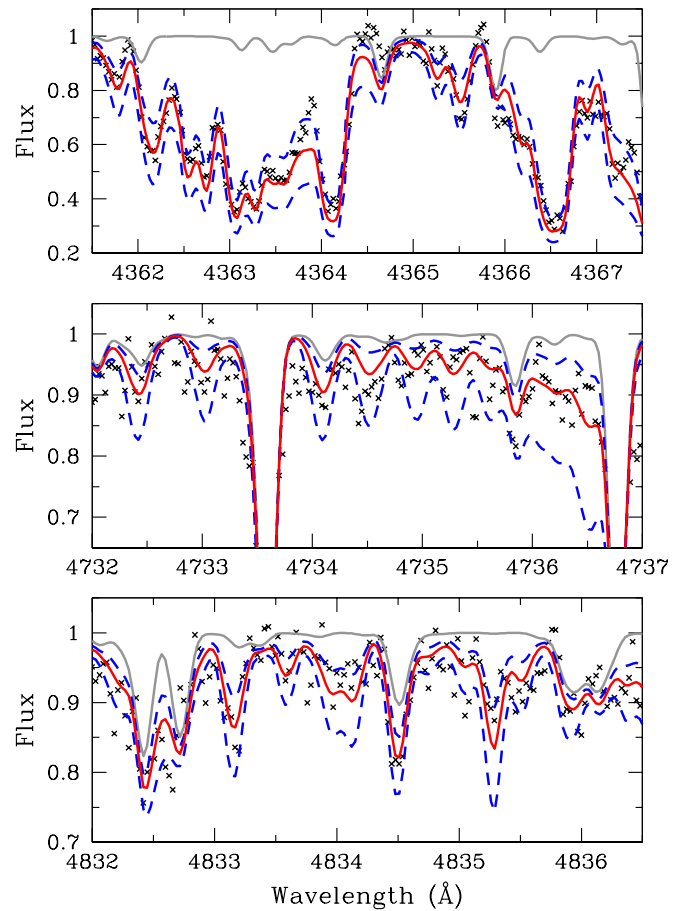}
      \caption{Spectra of different regions used for the measurement of carbon. Top panel shows part of the CH region at $\sim$4300\text{\AA} (G-band), middle panel shows the C$_2$ band, and lowest panel shows the CH band (A-X system) at $\sim$4850~\text{\AA}. In all cases, solid red lines show best fits of the regions (as listed in the text), dashed blue lines show carbon values $\pm$0.20 from that fit, and solid gray lines show how the spectrum would look without any carbon. Note that the scale on the y-axis is not the same for all panels. 
      }    
         \label{CH}
   \end{figure}
   
      \begin{figure}
   \centering
   \includegraphics[width=\hsize]{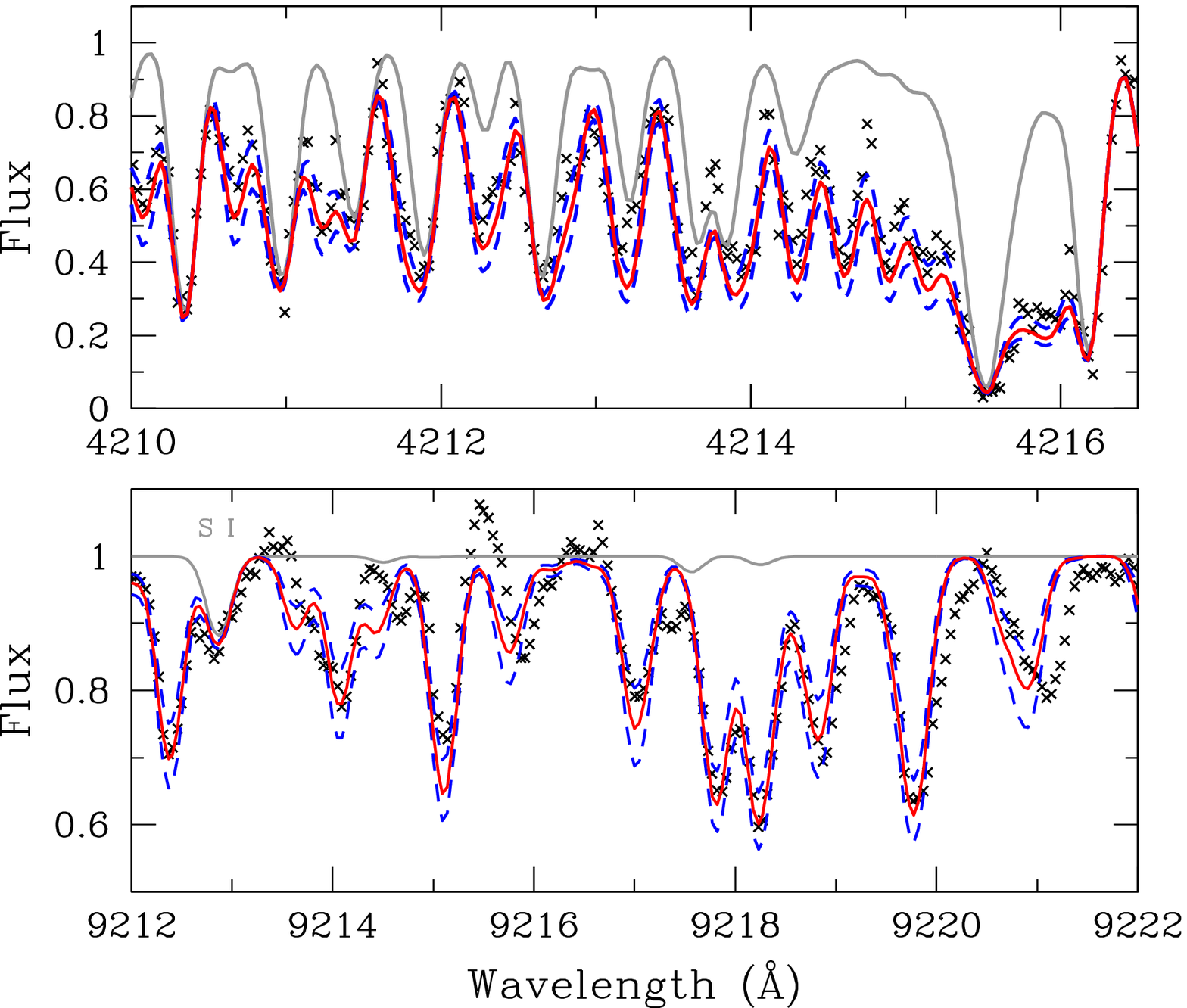}
      \caption{Upper panel shows strong CN molecular band at 4215~\text{\AA} (B-X system), while lower panel shows an example of the CN molecular band seen in the redder part of the spectrum. In both cases, solid red lines show the best fit of $\log \epsilon (\text{N})~=~7.07$, dashed blue lines show nitrogen values $\pm$0.20 from that fit, and solid gray lines what the spectrum would look like without any molecular CN lines. Note that the scale on the y-axis is not the same for both panels.
      }
         \label{CN}
   \end{figure}

\subsection{Oxygen, carbon, and nitrogen}

The oxygen abundance was derived from the forbidden [O~I] lines at 6300 and 6364~\AA , see Fig. \ref{O}. Both lines were slightly blended with weak CN molecular lines, but changing the adopted C and N values by $\pm$0.20 around the best fit, changed the measured oxygen abundances only by $\sim$0.05~dex. The abundance measurements of the lines agree within the errorbars, 
$\log \epsilon (\text{O})_{6300\text{\AA}}~=~7.30~\pm~0.14$ and $\log \epsilon (\text{O})_{6364\text{\AA}}~=~7.46~\pm~0.30$. The final value of $\log \epsilon (\text{O})~=~7.33\pm0.18$ was adopted by weighting the two [O~I] lines with their errors. 

The [O~I] lines of the triplet at 7774~\text{\AA} were too weak to be accurately measured, considering the noise and slight blending, and they were therefore not included in the measurement.

The carbon abundance for ET0097 was determined by fitting three different molecular bands in regions of 20~\AA. Parts of these bands are shown in Fig. \ref{CH}. In some wavelength ranges, the molecular lines were saturated and in others very weak. Therefore, we only used regions where the $\chi^2$ of the fit was sensitive to the assumed C abundance. The CH G-band at $\sim$4300~\text{\AA}, spreading throughout the range $\sim$4200-4400~\AA, resulted in $\log \epsilon (\text{C})_{\text{CH:}4300\text{\AA}}=6.98$. The C$_2$ band around $\sim$4700~\text{\AA} gave a result 0.20~dex lower, $\log \epsilon (\text{C})_{\text{C2:}4700\text{\AA}}=6.78$. A relatively weak CH band (A-X system) at $\sim$4850~\AA$ $ was also used to measure carbon, yielding a value 0.25~dex higher than the stronger CH-band, $\log \epsilon (\text{C})_{\text{CH:}4850\text{\AA}}=7.23$. The final measured value of $\log \epsilon (\text{C})=7.00\pm0.10$ was adopted by averaging both CH bands and the C$_2$ band, weighting them with the size of the region available for the measurements.

The CN band at 4215~\AA\ (B-X system) appeared clearly and without severe saturation (see Fig. \ref{CN}). A big portion of the observed spectrum for ET0097 was covered with CN lines, due to the relatively low temperature of the star and the high C and N abundances. The nitrogen was therefore measured using both the band at 4215~\text{\AA} and CN molecular lines in the wavelength range $\sim$6200-9400~\text{\AA} (A-X system), giving the final result of $\log \epsilon (\text{N})=7.07\pm0.20$. Regions where the $\chi^2$ of the fit was insensitive to the nitrogen abundance were excluded. The nitrogen measurements, which cover 104 regions of 20~\text{\AA}, show no trend with wavelength and are very consistent with a low scatter, $\sigma=0.05$. The measurements in the redder part of the spectrum and the CN band at 4215~\text{\AA} are in perfect agreement, see Fig.~\ref{CN}, indicating that the CN molecular parameters in the red are reliable. The uncertainty in the measurement from the CN band in the blue is higher than in the red, mostly due to the uncertain continuum determination. Using the larger wavelength range, however, the statistical errors for the nitrogen become negligible, and the real uncertainties come from errors in the carbon and oxygen measurements.

The B-X system of CN at 3888~\text{\AA} was extremely strong in this star and wiped out all continuum points until the bluest end of the spectrum at $\sim$3770~\text{\AA}. Thus, it was not included in the abundance measurement of nitrogen. However, as far as rough continuum estimation allowed, it was consistent with the result obtained using the other CN molecular bands. 

Because CO locks away a sizable fraction of the C available in the star, both CH bands and the C$_2$ band are sensitive to the oxygen value, and the nitrogen abundance measured from the CN molecular bands is sensitive both to the oxygen and the carbon abundances. Therefore, the oxygen-carbon-nitrogen measurements were iterated several times to minimize the influence of the errors of one element on the other elements as much as possible.

      \begin{figure}
   \centering
   \includegraphics[width=\hsize]{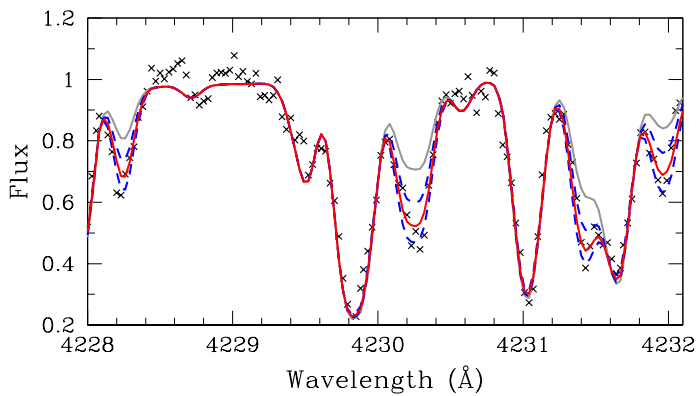}
      \caption{Example of the $^{13}$C features used for evaluating the $^{12}$C/$^{13}$C isotope ratio. Solid gray line shows the synthetic spectrum assuming  $\log^{12}\text{C}/^{13}\text{C}=1.3$, value typical for unmixed giants \citep{Spite2006}, solid red line shows $\log^{12}\text{C}/^{13}\text{C}=0.8$, and dashed blue lines show ratios $\pm$0.2 from that value.
      }
      
         \label{C13}
   \end{figure}

\subsection{Indicators of mixing, $^{\textsl{12}}$C/$^{\textsl{13}}$C ratio and Li}\label{sec:mixing}

When a star moves along the RGB, it undergoes a second episode of mixing (at the so-called RGB-bump) that lowers the carbon abundance at its surface, and increases the nitrogen. This upper RGB phase is reached at the luminosity $\log L_\star/L_\odot\sim2.2$, for stars in the metallicity range $-2.0~\lesssim~$[Fe/H]$~\lesssim-1.0$ \citep{Gratton2000} and when $\log L_\star/L_\odot\sim2.6$ for $\text{[Fe/H]}~\lesssim-2.5$ \citep{Spite2006}. Before an RGB star reaches this high luminosity, its C and N abundances do not change significantly during the star's evolution \citep{Gratton2000}. 

Thus, a low C value and high N abundance are usually good indicators of this second episode of mixing. With the high luminosity of $\log L_\star/L_\odot=3.1\pm0.1$, ET0097 has a high nitrogen abundance, $\text{[N/Fe]}=1.18\pm0.20$, but also high $\text{[C/Fe]}=0.51\pm0.10$, so if it has undergone this mixing, it would be inherently more carbon-rich.

During mixing, the most fragile element, Li, is destroyed in the deeper layers of the star, so low Li abundances indicate that mixing has occured. For main sequence (MS) stars, the Li abundance generally lies along the so-called Spite plateau, $\log \epsilon(\text{Li})\approx2.2$. When a star enters the lower RGB phase, the Li at the surface is diluted with material coming from the deeper layers, and the Li abundance stabilizes at an average value of $\log\epsilon(\text{Li}) \approx 1$ along the lower RGB. However, when the second episode of mixing occurs, practically all the remaining Li is destroyed, yielding a very low value of $\log\epsilon(\text{Li}) \leq 0$ \citep{Gratton2000}.

The lithium abundance for ET0097 was obtained by measuring the resonance doublet at 6707~\text{\AA}, yielding a value of $\log \epsilon(\text{Li})=-0.12$, but due to the weakness of the line an upper limit of $\log \epsilon(\text{Li})<0.17$ was adopted, consistent with most of the Li being destroyed.

Another signature of deep mixing is a low isotope ratio, $\log ^{12}$C/$^{13}$C < 1.0 \citep{Spite2006}. To measure $^{12}$C/$^{13}$C, the total carbon abundance was kept constant and the $^{13}$C lines  were fitted with synthetic spectra with different isotope ratios, see Fig. \ref{C13}. The ratio was determined to be $\log^{12}\text{C}/^{13}\text{C}=0.77\pm0.03$.

The main signatures of mixing in a high luminosity RGB star are thus present in ET0097, with $\log L_\star/L_\odot=3.1$, $\text{[N/Fe]}=1.18$, $\log^{12}\text{C}/^{13}\text{C}=0.77$ and a best fit of $\log \epsilon(\text{Li})~=~-0.12$. We therefore conclude that this star has undergone mixing, and that its carbon abundance was higher at the earlier evolutionary stages. 

By assuming most of the nitrogen present in the star was converted from carbon, we get an estimate of the original C abundance at the surface of the star: $\text{[C/Fe]}\approx0.8$. In the Galactic halo, mixing in RGB stars is observed to lower the surface C abundance on average by $\sim$0.5 \citep{Gratton2000,Spite2005}, with a scatter of $\sim$0.25~dex (\citealt{Spite2005}, for $-3\leq\text{[Fe/H]}\leq-2$), consistent with the correction made here. An online tool to correct for mixing has been provided by \citet{Placco2014}, which gives $\text{[C/Fe]}\approx0.9$ as the initial abundance for ET0097, while here we adopt the more conservative value of $\text{[C/Fe]}\approx0.8$. The estimate for the star's initial carbon enhancement thus falls under the classical definition of a CEMP ($\text{[C/Fe]}\geq0.7$) and the same conclusion is reached using the uncorrected value with the luminosity dependent definition of CEMP stars, from \citet{Aoki2007}.

\subsection{Alpha elements}

The main production sites of the alpha elements (made up of alpha particles) are normal core collapse Type~II Supernovae. Early in the star formation history of any galaxy, Type II SN are believed to be the main contributors of metals, so the early ISM holds the imprint of their yields, and stars formed at earlier times typically show an enhancement in [$\alpha$/Fe] (e.g., $\text{[Mg/Fe]}\gtrsim0.3$). About 1-2~Gyr after the first SN~II, Type~Ia Supernova start to contribute (e.g., \citealt{deBoer2012}), and so [$\alpha$/Fe] starts to decrease with increasing [Fe/H]. This so-called `knee', happens around $\text{[Fe/H]}\sim-1.7$ in Sculptor \citep{Tolstoy2009}.

In ET0097, the abundance measurements for Mg, Ca, and Ti were relatively straight forward, all having many lines that did not show any signs of blending from atomic or molecular lines, see Table \ref{table:linelist}. On the other hand, all lines for Si were either weak and/or slightly blended. Though the blending was accounted for in the synthetic spectra, it added to the uncertainty of the measurements. However, there were many Si lines available in the observed wavelength range, and the final abundance was measured from the seven~best lines. Some of these were weak and unblended, others slightly stronger but were blended. The number of lines and the reasonable scatter makes the result robust.

There were three S lines in the reddest part of the spectrum $\sim$9200~\text{\AA}, which were affected by reasonably strong CN lines, see the lower panel of Fig. \ref{CN}. Two S lines were relatively unblended, both of them yielding very similar results (agreeing within 0.05~dex). The line at 9237.5~\text{\AA} was not used, however, it was consistent with the value obtained for the other two lines. 

All alpha elements in ET0097 show an overabundance relative to iron, $\text{[$\alpha$/Fe]}>0$, similar to what is seen for other stars in Sculptor and in the Galactic halo at comparable iron abundances, $\text{[Fe/H]}\approx-2$.

\subsection{Odd-Z elements}

The abundances of Na, Al, and K were determined from resonance lines that are very sensitive to non-local thermodynamic effects (NLTE) effects. However, similar correction factors are expected for RGB stars with comparable metallicity, $T_\textsl{eff}$ and $\log g$, so when comparing abundances of ET0097 with similar stars in the Galactic halo or Sculptor, similar NLTE corrections can be applied. \citet{Cayrel2004} adopted a correction of $-0.50$~dex for Na in giants, $+0.65$~dex for Al and $-0.35$~dex for K. Here, the same lines were used for these elements, so similar corrections can be made to the LTE values listed in Table \ref{table:abundances}. More detailed NLTE calculations for stars with similar stellar parameters are presented in \citet{Andrievsky2007,Andrievsky2008,Andrievsky2010}.

The Na abundance was determined from the D resonance lines at 5890~\text{\AA} and 5896~\text{\AA}, and a weaker line at 8183.3~\text{\AA}. None of these lines showed any signs of blending and gave the weighted average of $\log \epsilon (\text{Na})=3.81\pm0.24$.

Two Al lines were visible in the spectrum, the resonance doublet at 3944~\text{\AA} and 3961.5~\text{\AA}. Both lines were heavily blended, in particular the line at 3944~\text{\AA}, so it was not included in the measurement. The result remains rather uncertain, because of the blending and the difficult continuum evaluation in this region, $\log \epsilon (\text{Al})=3.62\pm0.80$, which is consistent with the line at 3944~\text{\AA}.

The K abundance was determined from two strong lines at 7665~\text{\AA} and 7699~\text{\AA}. Both gave consistent results with a difference of only 0.02 dex. The adopted abundance is therefore $\log \epsilon (\text{K})=3.45\pm0.07$. 

Both Sc and V had many unblended lines available, giving $\log \epsilon (\text{Sc})=1.10\pm0.04$ and $\log \epsilon (\text{V})=2.01\pm0.04$.

\subsection {Iron-peak elements}
In general, Fe-peak elements are believed to be created in supernova explosions. The elements Cr, Mn, Co, and Ni all had many available lines in the observed wavelength range, making it possible to discard those that were blended with molecular lines.

Only one rather weak line was available for Cu, at 5782~\text{\AA}, in a region of the spectrum that was relatively free of molecular lines. The line barely showed any blending, giving $\log \epsilon (\text{Cu})=1.44\pm0.26$. Three lines were available for Zn. They only showed minor blending and agree well with each other, $\log \epsilon (\text{Zn})=2.68\pm0.23$.

\subsection{Heavy elements}

Abundances for three elements of the lighter $n$-capture elements were measured: Sr, Y and Zr. Two lines were observed for Sr, one very strong and blended Sr~II line at 4078~\AA, giving the result $\log \epsilon (\text{Sr})_{4078\text{\AA}}=1.44\pm0.34$, and a Sr~I line free of blending at 4607~\text{\AA}, $\log\epsilon(\text{Sr})_{4607\text{\AA}}=1.70\pm0.16$. The weighted average of the two yields 
$\text{[Sr/Fe]}=0.71\pm 0.20$. Yttrium had six lines in the wavelength range, all showing only minor blending and very little scatter, giving the final value $\text{[Y/Fe]}=0.34\pm0.06$. Five lines are used for the measurement of Zr, three of them slightly blended, giving the result $\text{[Zr/Fe]}=0.37\pm0.07$. The lighter neutron-capture elements in ET0097 therefore all show overabundance with respect to iron, $\text{[Sr,Y,Zr/Fe]}>0.3$.

The heavier $n$-capture elements Ba, La, Ce, and Nd, all had five or more lines available. Of those, La and Ce showed more scatter between lines, $\sigma\sim 0.30$~dex (compared to $\sigma\sim 0.15$ of Ba and Nd), which was to be expected since many of the measured lines for these elements were blended and/or weak. Three Sm lines were measured with a scatter between lines, $\sigma=0.23$, and the final result is $\text{[Sm/Fe]}=0.11\pm0.32$. Eu was difficult to measure in this star since the four lines available were all heavily blended. However, all lines agree reasonably well with each other, with a scatter between lines of $\sigma=0.23$, giving the rather low value $\text{[Eu/Fe]}=0.03\pm0.42$. No trace of the Eu line at 6645~\text{\AA} was seen, but an upper limit of $\text{[Eu/Fe]}<0.30$ was determined, which is consistent with the four detected lines. The Dy abundance was derived from two weak and blended lines at 3945~\AA$ $ and 4103~\AA, $\text{[Dy/Fe]}=-0.03\pm0.53$. This is consistent with the best fits of two other weak lines at 3984 and 4450~\text{\AA}. Only upper limits could be determined for those: $\text{[Dy/Fe]}_{3984\text{\AA}}<0.51$ and $\text{[Dy/Fe]}_{4450\text{\AA}}<0.23$. There were no detectable lines for the elements Pr, Gd, Tb, Er, and Pb, giving upper limits for these elements in the range $\text{[X/Fe]}\sim$0.2-1.0~dex, which excludes the possibility of extreme overabundances. 

With low abundances for both the main $s$-process elements ($\text{[Ba/Fe]}<0$) and main $r$-process elements ($\text{[Eu/Fe]}<0.5$), ET0097 classifies as a CEMP-no star.

\section{Error analysis}

To evaluate the statistical uncertainties in the abundance determination of a line, $\delta_{noise,i}$, the noise in line-free regions neighboring the line was measured. The error was then determined as when the $\chi^2$ of the fit became larger than that of the noise. Since the spectrum was dominated by molecular lines, line-free regions were not always available. Although the molecular bands were reasonably well fitted as a whole with the synthetic spectra, in some regions individual lines were not. In those cases, the typical deviation of the spectrum from the best synthetic fit was measured and included in the noise estimate.

The individual lines showed different degrees of blending, and for elements with fewer than five measured lines, this was accounted for by weighting the different measurements with their errors as follows:

\begin{equation}
\log \epsilon (X) = \frac{\sum\limits_{i=1}^{N_X} \log \epsilon (X)_i \cdot w_i}{\sum\limits_{i=1}^{N_X} w_i} 
\end{equation} 
The sum runs over $N_X$ lines and the weights of individual lines were defined as:
\begin{equation}
w_i=\frac{1}{\delta^2_\textsl{noise,i}}
\end{equation}
where $\delta_\textsl{noise,i}$ is the statistical uncertainty of the abundance measurement of line $i$. For elements with five or more lines, this was not necessary and normal averages were used to determine the final abundances.\\

The final error for elements with four or fewer measured lines was calculated as follows:

\begin{equation}
\delta_{\textsl{noise}} = \sqrt{\frac{N_X}{\sum_i w_i}} 
\end{equation}
For elements with five or more lines, the total error from the noise was defined from the scatter of the measurements:

\begin{equation}\label{stat}
\delta_{\textsl{noise}} = \frac{\sigma}{\sqrt{N_X}} 
\end{equation}

   \begin{figure}
   \centering
   \includegraphics[width=\hsize-0.5cm]{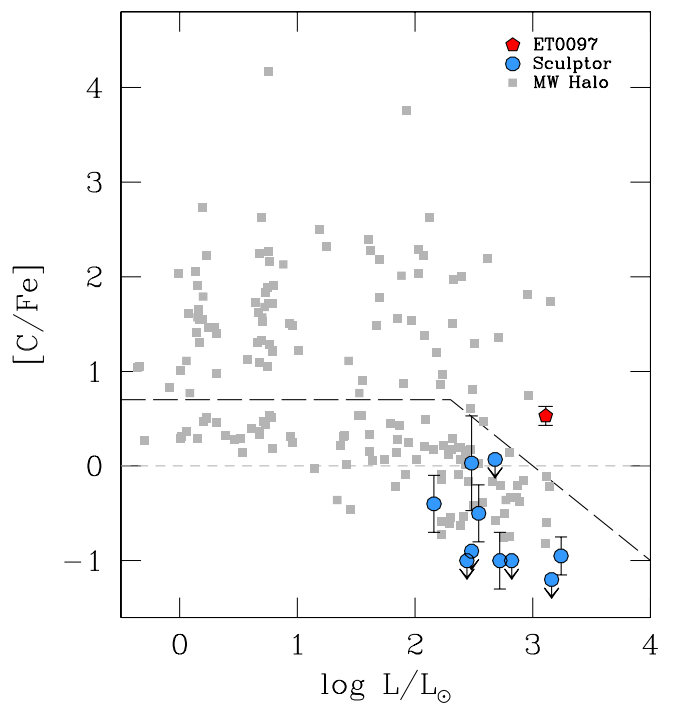}
      \caption{Carbon in Sculptor and the Galactic halo as a function of luminosity. Dashed line shows the definition of CEMP stars as proposed by \citet{Aoki2007}, to account for the mixing in stars at higher luminosities. Halo stars (gray squares) come from \citet{Yong2013a}, and Sculptor stars (blue circles) from: \citet{Tafelmeyer2010,Frebel2010,Kirby2012,Starkenburg2013}.}
      
         \label{fig:cempL}
   \end{figure}

Special care was taken in the evaluation of the errors on C and N abundances. These elements were measured over regions of 20~\text{\AA} and the final value was determined from the average of all measured regions. Measurement errors were calculated using Eq. (\ref{stat}). For C this gave an error of $\delta_C=0.05$ and $\delta_N=0.01$ for N. 

However, the CH and C$_2$ lines are sensitive to O values, so the effect of the oxygen error on these abundances was measured and included, yielding a total statistical error of $\delta_{\textsl{noise,C}}=0.09$. 

The CN molecular lines that were used to measure N are sensitive to both C and O abundances. Taking the effect of the uncertainties of these elements on the N measurements into account, the final error is $\delta_{\textsl{noise,N}}=0.20$.

The systematic errors coming from the uncertainties of the stellar parameters, $T_\textsl{eff}$, $\log g$ and $v_t$, were measured to be $\Delta\text{[Fe/H]}_\text{sp}=0.10$, and ranging from 0.02 to 0.06~dex for $\Delta$[X/Fe]$_\text{sp}$ (depending on the element). They were added quadratically to the $\delta_{\textsl{noise}}$ to obtain the adopted error, 
\begin{equation}\label{eq:totalerr}
\delta_{\textsl{total}}(\text{[X/Fe]})=\sqrt{ \delta_{\textsl{noise}}(\text{X})^2+\delta_{\textsl{noise}}(\text{Fe})^2+ \Delta\text{[X/Fe]}_\text{sp}^2} 
\end{equation}


\begin{longtab}
\begin{longtable}{l c c c r r r}
\caption{Carbon estimates and upper limits from CN molecular lines in the wavelength range 9100-9250~\text{\AA} (VLT/FLAMES). The estimates, $\text{[C/Fe]}_\text{est}$, are measured by assuming $\text{[C/N]}=-1.2$, the average value for mixed stars in the range $-3\leq\text{[Fe/H]}\leq-2$ \citep{Spite2005}. The upper limits, $\text{[C/Fe]}_\text{lim}$, are measured by assuming $\text{[N/Fe]}=0$, while all mixed stars in \citet{Spite2005} have $\text{[N/Fe]}>0.5$. This lower limit of N therefore gives an upper limit of C. In both cases a simple relation of oxygen with [Fe/H], as shown in Fig. \ref{fig:cplot}a, is assumed for stars that do not have known oxygen values. Iron values, luminosities, and oxygen measurements, when available, are obtained from Hill et al. in prep. The solar values are adopted from \citet{GrevesseSauval1998}.}  \\                                             
\hline              
Star	&	RA			&	DEC			&$\log L_\star/L_\odot$	&	[Fe/H]	&	[C/Fe]$_\text{est}$	&	[C/Fe]$_\text{lim}$	\\
\hline   
\endfirsthead
\caption{continued.}\\
 \hline                
Star	&	RA			&	DEC			&$\log L_\star/L_\odot$	&	[Fe/H]	&	[C/Fe]$_\text{est}$	&	[C/Fe]$_\text{lim}$	\\
\hline   
\endhead
ET0024	&	1	00	34.04	&	$-33$	39	04.6	&$	3.26	$&$	-1.24	$&$	-0.94	$&$	-0.74	$\\
ET0026	&	1	00	12.76	&	$-33$	41	16.0	&$	3.08	$&$	-1.80	$&$	-0.96	$&$	-0.74	$\\
ET0027	&	1	00	15.37	&	$-33$	39	06.2	&$	3.10	$&$	-1.50	$&$	-0.94	$&$	-0.70	$\\
ET0028	&	1	00	17.77	&	$-33$	35	59.7	&$	3.11	$&$	-1.22	$&$	-0.98	$&$	-0.78	$\\
ET0031	&	1	00	07.57	&	$-33$	37	03.9	&$	2.98	$&$	-1.68	$&$	-0.88	$&$	-0.58	$\\
ET0033	&	1	00	20.29	&	$-33$	35	34.5	&$	2.98	$&$	-1.77	$&$	-0.90	$&$	-0.60	$\\
ET0043	&	1	00	13.95	&	$-33$	36	39.2	&$	2.84	$&$	-1.24	$&$	-1.04	$&$	-0.88	$\\
ET0048	&	0	59	55.63	&	$-33$	33	24.6	&$	3.18	$&$	-1.90	$&$	-0.70	$&$	-0.24	$\\
ET0051	&	0	59	46.41	&	$-33$	41	23.5	&$	3.18	$&$	-0.92	$&$	-1.12	$&$	-1.08	$\\
ET0054	&	0	59	56.60	&	$-33$	36	41.7	&$	3.00	$&$	-1.81	$&$	-0.80	$&$	-0.44	$\\
ET0057	&	0	59	54.21	&	$-33$	40	27.2	&$	3.02	$&$	-1.33	$&$	-0.88	$&$	-0.62	$\\
ET0059	&	0	59	38.11	&	$-33$	35	08.0	&$	2.98	$&$	-1.53	$&$	-1.17	$&$	-1.13	$\\
ET0060	&	0	59	37.74	&	$-33$	36	00.0	&$	2.98	$&$	-1.56	$&$	-0.94	$&$	-0.72	$\\
ET0062	&	0	59	47.21	&	$-33$	33	36.9	&$	2.91	$&$	-2.27	$&$	<-0.50	$&$	<0.00	$\\
ET0063	&	0	59	37.22	&	$-33$	37	10.5	&$	2.99	$&$	-1.18	$&$	-0.96	$&$	-0.76	$\\
ET0064	&	0	59	41.40	&	$-33$	38	47.0	&$	2.96	$&$	-1.38	$&$	-0.88	$&$	-0.58	$\\
ET0066	&	1	00	03.60	&	$-33$	39	27.1	&$	2.94	$&$	-1.30	$&$	-1.00	$&$	-0.84	$\\
ET0067	&	0	59	37.00	&	$-33$	30	28.4	&$	2.90	$&$	-1.65	$&$	-0.96	$&$	-0.72	$\\
ET0069	&	0	59	40.46	&	$-33$	35	53.8	&$	2.86	$&$	-2.11	$&$	-0.82	$&$	-0.44	$\\
ET0071	&	0	59	58.27	&	$-33$	41	08.7	&$	2.92	$&$	-1.35	$&$	-0.98	$&$	-0.78	$\\
ET0073	&	0	59	53.99	&	$-33$	37	42.1	&$	2.82	$&$	-1.53	$&$	-0.90	$&$	-0.62	$\\
ET0083	&	0	59	11.83	&	$-33$	41	25.3	&$	2.97	$&$	-1.97	$&$	-0.76	$&$	-0.34	$\\
ET0094	&	0	59	20.65	&	$-33$	48	56.6	&$	3.18	$&$	-1.86	$&$	-0.82	$&$	-0.48	$\\
ET0095	&	0	59	20.80	&	$-33$	44	04.8	&$	3.09	$&$	-2.16	$&$	-0.80	$&$	-0.42	$\\
ET0103	&	0	59	18.85	&	$-33$	42	17.3	&$	2.95	$&$	-1.21	$&$	-1.04	$&$	-0.90	$\\
ET0104	&	0	59	15.14	&	$-33$	42	54.6	&$	2.90	$&$	-1.62	$&$	-0.82	$&$	-0.46	$\\
ET0109	&	0	59	28.29	&	$-33$	42	07.2	&$	3.24	$&$	-1.85	$&$	-0.76	$&$	-0.40	$\\
ET0112	&	0	59	52.27	&	$-33$	44	54.8	&$	3.11	$&$	-2.04	$&$	-0.72	$&$	-0.28	$\\
ET0113	&	0	59	55.68	&	$-33$	46	40.1	&$	3.08	$&$	-2.18	$&$	-0.74	$&$	-0.30	$\\
ET0121	&	1	00	00.49	&	$-33$	49	35.8	&$	2.94	$&$	-2.35	$&$	-0.83	$&$	-0.43	$\\
ET0126	&	0	59	42.57	&	$-33$	42	18.1	&$	2.99	$&$	-1.11	$&$	-0.96	$&$	-0.76	$\\
ET0132	&	0	59	58.24	&	$-33$	45	50.8	&$	2.88	$&$	-1.50	$&$	-0.88	$&$	-0.60	$\\
ET0133	&	0	59	47.67	&	$-33$	47	29.5	&$	2.88	$&$	-1.07	$&$	-1.04	$&$	-0.92	$\\
ET0137	&	1	00	25.30	&	$-33$	50	50.8	&$	3.27	$&$	-0.89	$&$	-0.98	$&$	-0.82	$\\
ET0138	&	1	00	38.12	&	$-33$	48	16.9	&$	3.12	$&$	-1.70	$&$	-0.92	$&$	-0.68	$\\
ET0139	&	1	00	42.50	&	$-33$	44	23.5	&$	3.18	$&$	-1.41	$&$	-0.96	$&$	-0.74	$\\
ET0141	&	1	00	23.84	&	$-33$	42	17.4	&$	3.08	$&$	-1.68	$&$	-0.82	$&$	-0.46	$\\
ET0145	&	1	00	20.75	&	$-33$	47	11.1	&$	3.00	$&$	-1.51	$&$	-1.26	$&$	-1.30	$\\
ET0147	&	1	00	44.27	&	$-33$	49	18.8	&$	3.03	$&$	-1.15	$&$	-1.24	$&$	-1.28	$\\
ET0150	&	1	00	22.98	&	$-33$	43	02.2	&$	3.05	$&$	-0.93	$&$	-1.18	$&$	-1.16	$\\
ET0151	&	1	00	18.29	&	$-33$	42	12.2	&$	2.98	$&$	-1.77	$&$	-0.86	$&$	-0.54	$\\
ET0158	&	1	00	18.96	&	$-33$	45	14.8	&$	2.85	$&$	-1.80	$&$	-0.96	$&$	-0.74	$\\
ET0160	&	1	00	22.33	&	$-33$	50	24.0	&$	2.90	$&$	-1.16	$&$	-1.00	$&$	-0.82	$\\
ET0163	&	1	00	24.63	&	$-33$	44	28.9	&$	2.82	$&$	-1.86	$&$	<-0.70	$&$	<-0.40	$\\
ET0164	&	1	00	33.86	&	$-33$	44	54.4	&$	2.82	$&$	-1.89	$&$	-1.08	$&$	-0.96	$\\
ET0165	&	1	00	11.79	&	$-33$	42	16.9	&$	2.88	$&$	-1.10	$&$	-0.92	$&$	-0.70	$\\
ET0166	&	1	00	10.49	&	$-33$	49	36.9	&$	2.83	$&$	-1.49	$&$	-0.90	$&$	-0.62	$\\
ET0168	&	1	00	34.32	&	$-33$	49	52.9	&$	2.83	$&$	-1.10	$&$	-1.10	$&$	-1.00	$\\
ET0173	&	1	00	50.87	&	$-33$	45	05.2	&$	3.23	$&$	-1.47	$&$	-0.84	$&$	-0.54	$\\
ET0198	&	1	00	09.18	&	$-33$	36	09.4	&$	2.78	$&$	-1.16	$&$	-1.11	$&$	-1.05	$\\
ET0200	&	1	00	14.81	&	$-33$	36	49.9	&$	2.77	$&$	-1.49	$&$	-0.96	$&$	-0.74	$\\
ET0202	&	1	00	21.08	&	$-33$	33	46.4	&$	2.72	$&$	-1.32	$&$	-1.04	$&$	-0.88	$\\
ET0206	&	1	00	10.38	&	$-33$	41	05.0	&$	2.72	$&$	-1.33	$&$	-0.98	$&$	-0.78	$\\
ET0232	&	0	59	54.47	&	$-33$	37	53.4	&$	2.80	$&$	-1.00	$&$	-1.19	$&$	-1.17	$\\
ET0236	&	0	59	30.44	&	$-33$	36	05.0	&$	2.74	$&$	-2.41	$&$	<-0.30	$&$	<0.40	$\\
ET0237	&	0	59	50.78	&	$-33$	31	47.1	&$	2.75	$&$	-1.61	$&$	-0.90	$&$	-0.62	$\\
ET0238	&	0	59	57.60	&	$-33$	38	32.5	&$	2.78	$&$	-1.57	$&$	-1.00	$&$	-0.80	$\\
ET0239	&	0	59	30.49	&	$-33$	39	04.0	&$	2.71	$&$	-2.26	$&$	<-0.40	$&$	<0.30	$\\
ET0240	&	0	59	58.31	&	$-33$	34	40.4	&$	2.77	$&$	-1.15	$&$	-1.02	$&$	-0.84	$\\
ET0241	&	1	00	02.69	&	$-33$	30	25.3	&$	2.79	$&$	-1.41	$&$	-0.94	$&$	-0.70	$\\
ET0242	&	1	00	02.23	&	$-33$	40	21.1	&$	2.86	$&$	-1.32	$&$	-0.94	$&$	-0.70	$\\
ET0244	&	0	59	59.65	&	$-33$	39	31.9	&$	2.73	$&$	-1.24	$&$	-1.04	$&$	-0.90	$\\
ET0275	&	0	59	15.13	&	$-33$	39	43.8	&$	2.70	$&$	-1.21	$&$	-1.08	$&$	-0.98	$\\
ET0299	&	0	59	08.60	&	$-33$	42	29.4	&$	2.70	$&$	-1.83	$&$	-0.66	$&$	-0.14	$\\
ET0300	&	0	59	22.12	&	$-33$	49	03.7	&$	2.75	$&$	-1.39	$&$	-0.98	$&$	-0.78	$\\
ET0317	&	0	59	49.91	&	$-33$	44	05.0	&$	2.81	$&$	-1.69	$&$	-0.96	$&$	-0.74	$\\
ET0320	&	0	59	45.31	&	$-33$	43	53.8	&$	2.76	$&$	-1.71	$&$	<-0.90	$&$	<-0.40	$\\
ET0321	&	1	00	06.98	&	$-33$	47	09.7	&$	2.78	$&$	-1.93	$&$	-1.10	$&$	-1.00	$\\
ET0322	&	1	00	05.93	&	$-33$	45	56.5	&$	2.74	$&$	-2.04	$&$	<-0.60	$&$	<0.00	$\\
ET0327	&	0	59	37.56	&	$-33$	43	33.5	&$	2.76	$&$	-1.32	$&$	-0.94	$&$	-0.68	$\\
ET0330	&	1	00	04.16	&	$-33$	43	32.4	&$	2.68	$&$	-2.00	$&$	-0.68	$&$	-0.18	$\\
ET0339	&	0	59	44.90	&	$-33$	44	35.1	&$	2.72	$&$	-1.08	$&$	-1.19	$&$	-1.17	$\\
ET0342	&	0	59	35.02	&	$-33$	50	55.9	&$	2.62	$&$	-1.35	$&$	-1.11	$&$	-1.03	$\\
ET0350	&	0	59	41.95	&	$-33$	45	03.7	&$	2.56	$&$	-1.90	$&$	<-0.50	$&$	<0.10	$\\
ET0354	&	0	59	55.87	&	$-33$	45	43.7	&$	2.56	$&$	-1.07	$&$	-1.15	$&$	-1.09	$\\
ET0363	&	0	59	53.08	&	$-33$	43	58.5	&$	2.52	$&$	-1.28	$&$	-1.06	$&$	-0.94	$\\
ET0369	&	1	00	11.73	&	$-33$	44	50.4	&$	2.80	$&$	-2.35	$&$	-0.81	$&$	-0.43	$\\
ET0373	&	1	00	17.36	&	$-33$	43	59.6	&$	2.74	$&$	-1.96	$&$	-0.88	$&$	-0.56	$\\
ET0376	&	1	00	15.18	&	$-33$	43	11.0	&$	2.78	$&$	-1.17	$&$	-0.96	$&$	-0.74	$\\
ET0378	&	1	00	21.17	&	$-33$	46	01.3	&$	2.77	$&$	-1.18	$&$	-0.96	$&$	-0.76	$\\
ET0379	&	1	00	14.58	&	$-33$	47	11.6	&$	2.72	$&$	-1.65	$&$	-1.11	$&$	-1.03	$\\
ET0382	&	1	00	17.60	&	$-33$	46	55.2	&$	2.72	$&$	-1.74	$&$	-1.04	$&$	-0.86	$\\
ET0384	&	1	00	26.29	&	$-33$	44	45.7	&$	2.71	$&$	-1.46	$&$	-1.10	$&$	-1.00	$\\
ET0389	&	1	00	12.52	&	$-33$	43	01.3	&$	2.68	$&$	-1.60	$&$	-0.98	$&$	-0.78	$\\
ET0392	&	1	00	25.04	&	$-33$	42	28.1	&$	2.67	$&$	-1.48	$&$	-0.86	$&$	-0.52	$\\
\hline 
\label{table:CNdata}    
\end{longtable}
\end{longtab}

%
\section{Results}
All element abundances for ET0097 are listed in Table \ref{table:abundances}.

\subsection{Carbon in Sculptor}

Carbon measurements have been previously attempted only for a limited number of stars in Sculptor, and the available measurements are shown in Fig. \ref{fig:cempL}. The dashed line shows the definition of CEMP stars, as proposed by \citet{Aoki2007}, with a slope to account for mixing in RGB stars at higher luminosities, which decreases the carbon abundance and increases the nitrogen at the surface of the star \citep{Gratton2000,Spite2006}. The high [N/Fe], low $\log \epsilon(\text{Li)}$, and low $^{12}\text{C}/^{13}\text{C}$ values show that ET0097 has undergone mixing, and was even more carbon-rich in the past, having $\text{[C/Fe]}\approx0.8$ (see Section \ref{sec:mixing} for details).

With a measured $\text{[C/Fe]}=0.51\pm0.10$, ET0097 is the only known star in Sculptor that falls under the definition of a CEMP star, and it seems to be around $\sim$1.5 dex higher in carbon than other stars of similar luminosity.

\begin{figure}
   \centering
   \includegraphics[width=\hsize-0cm]{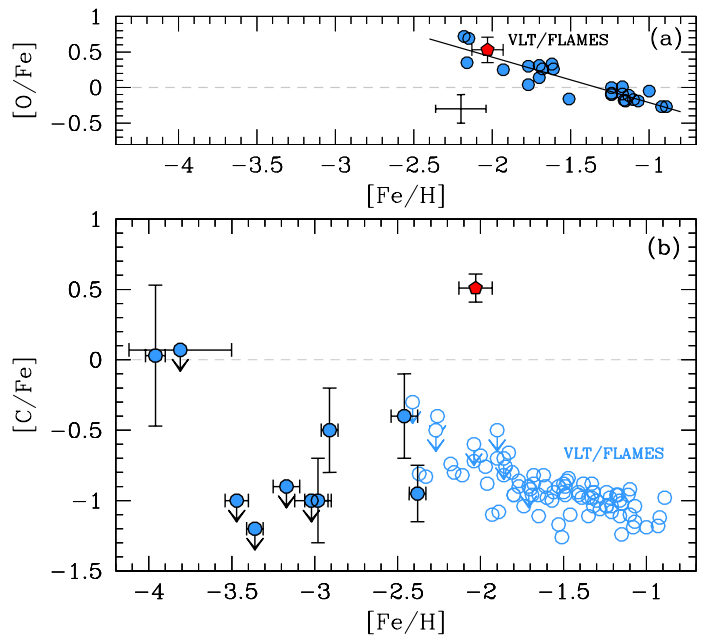}
      \caption{Same symbols are used as in Fig. \ref{fig:cempL}. \textbf{(a)} Oxygen in Sculptor as a function of [Fe/H]. Sculptor stars (blue filled circles) come from Hill et al. in prep. Representative error bar for the measurements is shown; \textbf{(b)} Carbon in Sculptor as a function of [Fe/H]. Solid points are direct carbon measurements of the stars shown in Fig. \ref{fig:cempL}. Open circles show estimates of C abundances for 85 stars (VLT/FLAMES spectra) from CN molecular lines in the wavelength range 9100-9250~\AA. Ratio typical for mixed stars, $\text{[C/N]}=-1.2$ \citep{Spite2005}, is assumed, and for stars with unknown oxygen abundance, a simple trend of [O/Fe] with [Fe/H] is adopted, as shown with a line in~(a). Two stars from \citealt{Starkenburg2013} that have the same [C/Fe] at $\text{[Fe/H]}\sim-3$ are moved by 0.02~dex to each side, to be visible as two stars. Same is done for two stars with similar C estimates at $\text{[Fe/H]}=-2.35$.}
         \label{fig:cplot}
   \end{figure}

The same stars are shown as a function of iron abundance in Fig. \ref{fig:cplot}b. The most carbon-rich star in the sample, ET0097, is also the most iron-rich, while the fraction of carbon-rich stars increases with decreasing metallicity in the Galactic halo (e.g., \citealt{Lee2013} and references therein). Unlike the other Sculptor stars in Fig. \ref{fig:cempL} and \ref{fig:cplot}b, ET0097 was not chosen for closer observation based on low [Fe/H]. The UVES spectrum for ET0097 was taken after the C-enhancement was discovered from strong CN molecular lines around 9100-9250~\text{\AA}. More than 80 other stars were also observed in this wavelength range (Sk\'{u}lad\'{o}ttir et al. in prep.), and all but the most metal-poor stars have a clear detection of the molecular lines, but only ET0097 stands out, having exceptionally strong CN lines.

To use these CN lines to estimate the C abundances in this sample, some assumptions need to be made about the oxygen and nitrogen. All the stars in this sample are within the central $25^\prime$ diameter region of Sculptor and bright enough to ensure reasonable signal-to-noise at high spectral resolution. These high luminosity RGB stars ($\log L_\star/L_\odot>2.5$ for all stars, see Table \ref{table:CNdata}) are expected to have undergone similar mixing to ET0097, decreasing the C at the surface and increasing the N abundance. Therefore, a typical value for mixed stars, $\text{[C/N]}=-1.2$ \citep{Spite2005}, is adopted here for the sample. Some of the stars already have measured O abundances and show a simple trend with iron, see Fig. \ref{fig:cplot}a. The same trend with [Fe/H] is therefore assumed for stars with unknown O abundances. 

The C~estimates of 85 stars, calculated with these assumptions, are also shown in Fig. \ref{fig:cplot}b (and listed in Table~\ref{table:CNdata}). None of these stars show any sign of being carbon-enhanced, even if corrected for internal mixing, which has been observed to lower the [C/Fe] abundance on the surface of stars by $\sim$0.5~dex \citep{Gratton2000,Spite2005}.

A different approach can be applied. In the sample of \citet{Spite2005}, all mixed stars have $\text{[N/Fe]}>0.5$, so by assuming $\text{[N/Fe]}=0$, which is a very conservative lower limit for nitrogen in these stars, we are also able to obtain upper limits for [C/Fe] from the Sculptor CN measurements (assuming the same oxygen values as before), see Table~\ref{table:CNdata}. If there are any unmixed stars in our sample, then $\text{[N/Fe]}=0$ is a reasonable abundance estimate for these stars \citep{Spite2005}. Using these assumptions, all 85 stars have $\text{[C/Fe]}_\text{lim}\leq0.4$. 

So with the exception of ET0097, which clearly stands out from the rest, no other star (mixed or unmixed) in this sample is likely to be inherently carbon-enhanced. In particular, by combining these estimates with the literature data, ET0097 is the only CEMP star from the sample of 22 stars in Sculptor with $\text{[Fe/H]}\leq-2$.

  \begin{figure*}
   \centering
   \includegraphics[width=\hsize-0.5cm]{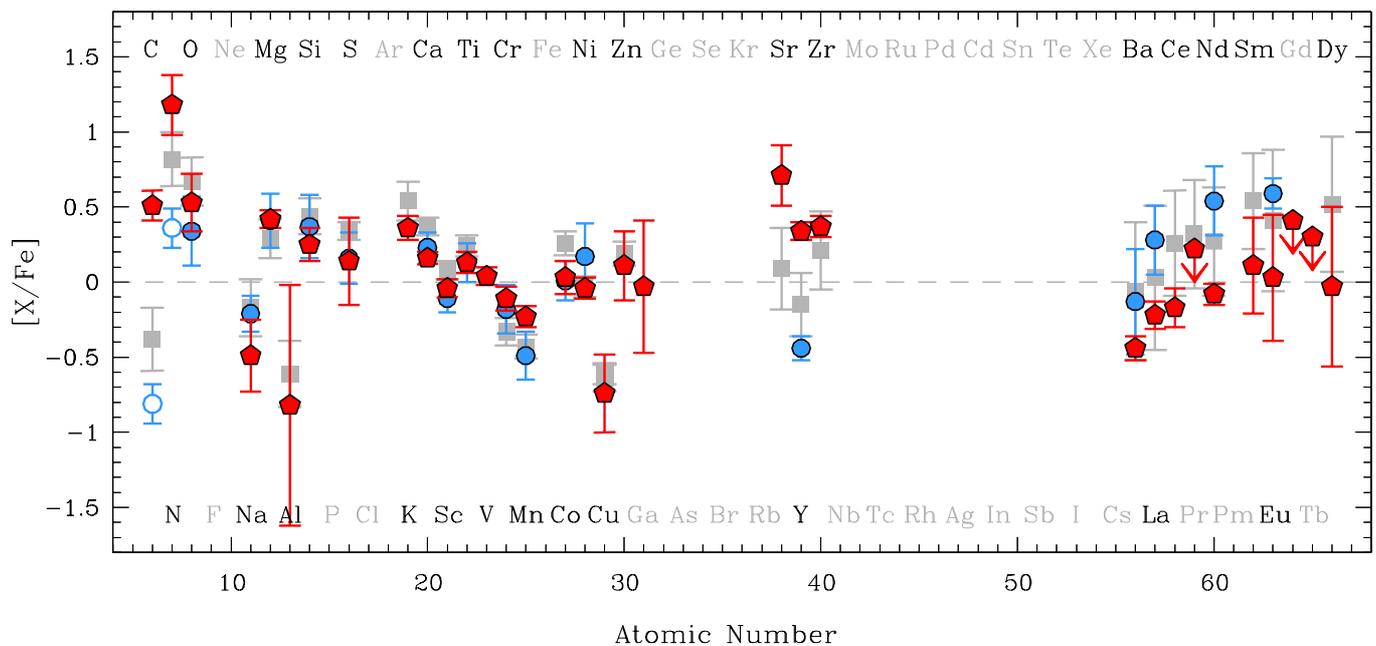}
      \caption{Element abundances for ET0097 (red pentagons), Sculptor (blue circles) and the Galactic halo (gray squares). Sculptor abundances are averaged over all stars with $-2.35\leq\text{[Fe/H]}\leq-1.75$. For Sculptor, most elements come from Hill et al. in prep. Exceptions are C and N estimated here from CN molecular lines (see Fig. \ref{fig:cplot} for detail), and S which comes from \skulad\ The Galactic halo abundances are averaged over stars with $-3.0\leq\text{[Fe/H]}\leq-2.0$. Abundance averages for C and N in the Galactic halo are taken from mixed, C-normal stars in \citet{Spite2005}. Other elements up to Zn are taken from \citet{Cayrel2004}, with the exceptions of S, which is from \citet{Spite2011}, and Cu, which is taken from both halo and disk stars in \citet{Mishenina2002}. Elements heavier than Zn come from \citet{Francois2007}. Error bars for Sculptor and the Galactic halo represent 1$\sigma$ of the scatter of [X/Fe] over each sample. No NLTE corrections have been applied, but are expected to be similar for all stars.
      }
         \label{fig:ele}
   \end{figure*}

\subsection{The general abundance pattern}

The abundance pattern of ET0097 is compared to what is seen in the Galactic halo and other stars in Sculptor, in Fig.~\ref{fig:ele}. Note that none of the abundances have been corrected for NLTE effects. For many elements this correction can be significant, in particular for Na, Al, and K, reaching up to $\sim$0.6 dex. However, the goal here is not to study the trends for these elements. Since both the Galactic halo and Sculptor samples consist of giant stars with similar $T_{\textsl{eff}}$, $\log g$ and [Fe/H] as ET0097, and have the same measured lines for these elements, any NLTE corrections are expected to be similar for all stars.

ET0097 is the only star in Sculptor in the metallicity range $-2.35\leq$[Fe/H]$\leq-1.75$ to have measured carbon and nitrogen, and, in fact, it has the only known nitrogen abundance in this galaxy. To estimate the C and N in stars of similar metallicity, we use the CN molecular lines in the region 9100-9250~\text{\AA}, from VLT/FLAMES data, see the previous section for details.

Compared to those estimates, ET0097 seems to be enhanced both in C and N with respect to other stars in Sculptor of similar [Fe/H], where the difference in carbon is $\gtrsim$1~dex, and  $\gtrsim$0.5~dex in nitrogen. Adopting a different [C/N] ratio could increase the N abundance estimate, bringing it closer to ET0097, but that would naturally decrease the C, making the difference there even bigger. Comparing ET0097 to carbon-normal, mixed RGB stars in the Galactic halo \citep{Spite2005}, the nitrogen seems to be rather high, but it does not stand out significantly from the scatter. The C in this star is however clearly enhanced compared to similar stars in the Galactic halo. Finally, we note that the [C+N/Fe] in the Sculptor sample seems to be lower than what is observed in the Galactic halo.

In Fig. \ref{fig:ele}, it is clear that when other elements up to Zn are compared with the mean for Sculptor and the Galactic halo, ET0097 does not stand out significantly in any way, and falls within the scatter of stars with similar metallicity. However, ET0097 does show a different pattern in elements heavier than Zn. The lighter neutron-capture elements (sometimes called weak $r$-process elements), Sr, Y, and Zr, are enhanced compared to what is typical in Sculptor, while the heavier $n$-capture elements are depleted, or at the lower end of the trend (Ba). The Galactic halo shows a very large scatter of the $n$-capture elements in the metallicity range $-3.0\leq$[Fe/H]$\leq-2.0$, so both ET0097 and other stars in Sculptor fall within the scatter seen in these elements,with the exception of Sr and Y, which appear above the observed scatter. 

\begin{figure}
   \centering
   \includegraphics[width=\hsize-0.5cm]{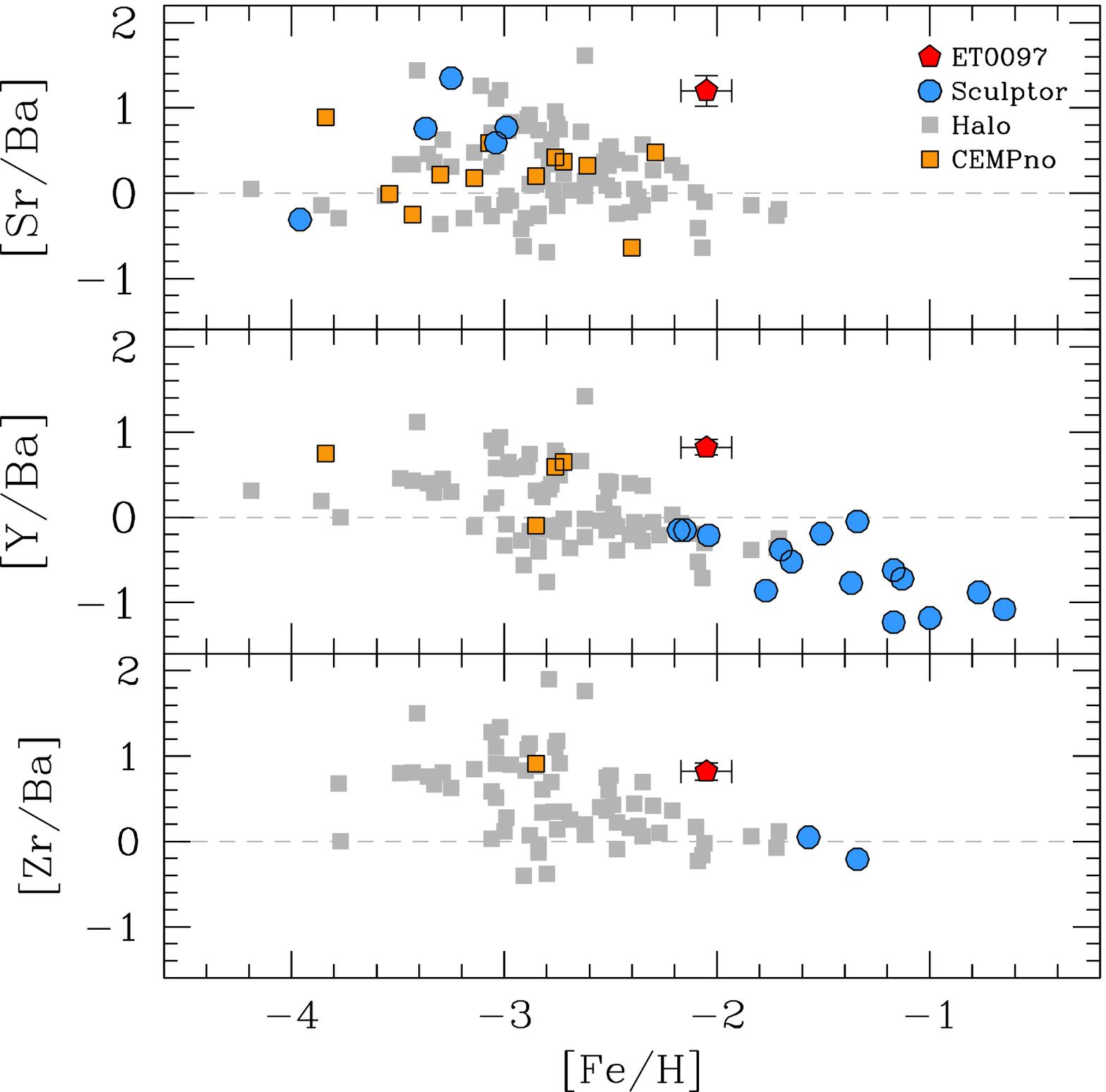}
      \caption{Light neutron-capture elements, Sr, Y, Zr, with respect to Ba, as a function of Fe abundance. Sculptor stars are from Hill et al. in prep.; \citet{Shetrone2003,Geisler2005,Tafelmeyer2010,Starkenburg2013}. Galactic halo stars are taken from: \citet{JohnsonBolte2002,Honda2004b,Aoki2005,Francois2007}. CEMP-no stars in the halo are taken from a compilation by \citet{Allen2012}, including data from: \citet{Norris2001,Norris2002,Giridhar2001,PrestonSneden2001,Aoki2002a,Aoki2002b,Aoki2002c,Aoki2004,Aoki2007,Depagne2002,Honda2004b,Barklem2005,Cohen2006,Cohen2008,Sivarani2006}.     
      }
         \label{fig:light}
   \end{figure}

\begin{figure}
   \centering
   \includegraphics[width=\hsize-0.5cm]{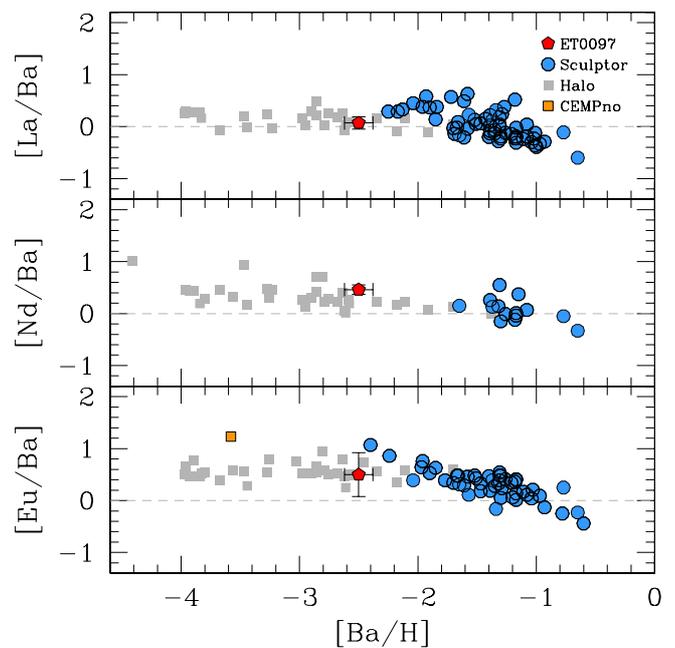}
      \caption{The relative abundances of the heavier neutron-capture elements. References are the same as in Fig. \ref{fig:light}.}
         \label{fig:heavy}
   \end{figure}

Since ET0097 shows high abundances of light $n$-capture elements and low abundances of the heavier $n$-capture elements, naturally the abundance ratios [Sr,Y,Zr/Ba] are high, see Fig. \ref{fig:light}. Here ET0097 is clearly different from the trend seen in the Galactic halo and in Sculptor. Similar abundance ratios are certainly seen in the halo (e.g., \citealt{Honda2004b,Francois2007}), but typically at lower iron abundance ($\text{[Fe/H]}\lesssim-3$). This result is not limited to a comparison with Ba, ET0097 still stands out when any of the other heavier $n$-capture elements are used as a reference element. In fact, ET0097 shows the same relation to the heavier $n$-capture elements as seen in the Galactic halo, see Fig.~\ref{fig:heavy}.

%

\section{Origin of the abundance pattern}

\subsection{Alpha and iron-peak elements}
In ET0097, alpha and Fe-peak elements from O to Zn show abundances comparable to what is seen both in the Galactic halo and in Sculptor for stars with similar [Fe/H], (see Fig.~\ref{fig:ele}). The most probable explanation is that the bulk of these elements comes from similar sources, such as low-metallicity Type~II supernovae of 11-40~M$_\odot$ \citep{WoosleyWeaver1995}, which are believed to be the main producers of these elements in the early universe, or possibly massive zero-metallicity SN of 10-100~M$_\odot$ \citep{HegerWoosley2010}, which have been shown to predict similar abundances as seen in the Galactic halo, hence comparable with ET0097.

\subsection{Carbon-enhancement}
The origin of the carbon enhancement in CEMP-no stars is still debated and a variety of processes have been invoked (see, e.g., \citealt{Norris2013}). 

In some of these scenarios, the carbon enhancement is explained by mass transfer from a companion. Three different HR velocity measurements for ET0097 were obtained in Hill et al. (in prep.), Sk\'{u}lad\'{o}ttir et al. (in prep.), and this work. A comparison was made between the two other studies for the 86 stars they had in common. ET0097 showed similar scatter in its velocity measurements compared to other stars in Sculptor. On average the difference between the samples was 1.5~km/s, and ET0097 had a difference of 1.4~km/s in the two measurements. Therefore, close binarity that favors mass transfer seems unlikely, though it cannot be completely excluded with the present data. 

However, even if this star does have a companion or did at some point, the abundance pattern is not easily explained with binarity. Mass transfer from an AGB-companion as seen in CEMP-$s$ stars can be excluded because $\text{[Y/Ba]}<0$ is expected \citep{Travaglio2004}, which is clearly not consistent with ET0097 (see Fig. \ref{fig:light}). Mass transfer from rapidly rotating stars, which are known to produce a lot of C, is also a possible source for the carbon enhancement. But from these stars strong enhancements in N and O, comparable to the C enhancement, are also expected \citep{Meynet2006}, and this is not consistent with the observations, which in particular show no enhancement in oxygen compared to other Sculptor and halo stars. This scenario can therefore be excluded.

Another possible scenario is that CEMP-no stars formed out of gas that has been enriched by faint supernovae with mixing and fallback which produce significant amounts of C but minimal Fe. Apart from the excess of carbon, faint SN show a general abundance pattern that is very different from normal Type II SN, with an excess of N and O. The ejecta from these stars are also predicted to have a very pronounced odd-even effect among iron-peak elements, showing up as very low abundance ratios, e.g., [V/Fe] and [Mn/Fe], which are not compatible with the results presented here. 

However, by assuming ET0097 formed out of gas containing a mixture of yields from faint SN and normal SN, it is possible to explain the carbon enhancement in this star. By assuming that normal SN Type~II enrich the gas up to $\text{[Fe/H]}\approx-2$, and that the gas was already pre-enriched with faint SN yields, as presented by \citet{Iwamoto2005} to match the hyper metal-poor star HE0107-5240, we require that the fraction of faint SN to normal SN is such that the gas reaches $\text{[C/Fe]}=0.8$ (a reasonable assumption for the initial value for ET0097). Though these faint SN yields also show enhancements in N and O, they are considerably smaller than for C. Therefore, the addition of these elements in ET0097 from faint SN would be undetectable in this mixture of yields, falling well within the error of the measurements. The effects on the abundances of other elements in the star are even less pronounced. All the peculiarities of the faint SN yields are swamped by the SN Type~II enrichment, leaving the high [C/Fe] value as the only evidence of its contribution. Therefore, it is indeed possible that the carbon enhancement in ET0097 is the result of (partial) enrichment with the products of faint SN. This is discussed in more detail in Section \ref{sec:PFS}.

\subsection{The lighter $n$-capture elements}

The main $r$- and $s$-process are excluded as dominant sources of Sr, Y, and Zr in ET0097 because they are predicted to produce much higher abundance ratios of heavy to light neutron-capture elements than are consistent with the data. The enhancements of Sr, Y, and Zr, most probably come from the weak $r$-process (e.g., \citealt{ArconesMontes2011}), which is predicted to produce significant amount of these elements, but minimal heavier $n$-capture elements ($Z\geq$56). Another possibility for the source of these elements is the weak $s$-process that occurs in fast-rotating massive zero-metallicity stars. Models of these stars have been able to reproduce the scatter of these elements observed in the Galactic halo \citep{Cescutti2014}.  However, the data presented here are not sufficient to distinguish between the different possible scenarios, and models that predict excesses of Sr, Y, and Zr, without significant effects on other element abundances, are consistent with observations of ET0097.  

\begin{figure}
   \centering
   \includegraphics[width=\hsize-0cm]{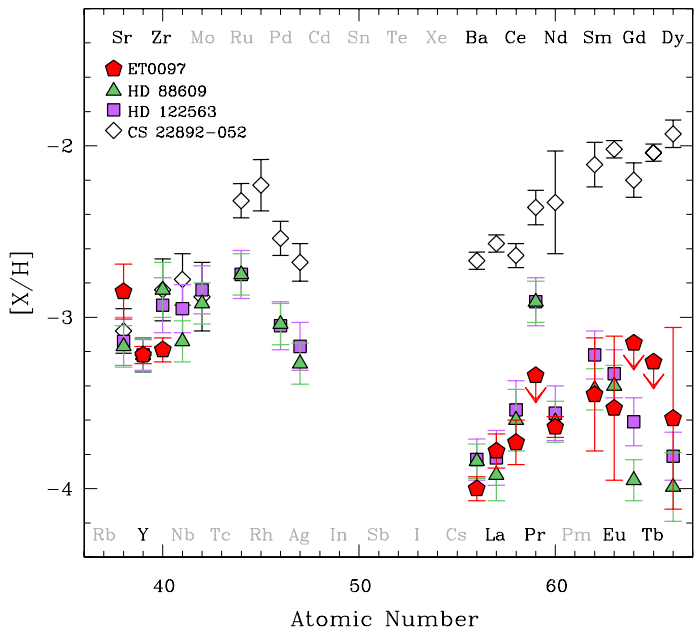}
      \caption{The $n$-capture elements of ET0097 compared to two stars with weak $r$-process enhancements, HD~88609 and HD~122563 \citep{Honda2007}, and one main $r$-process rich star, CS~22892-052 \citep{Sneden2003}. All four stars are normalized to the same [Y/H] value as HD~88609.}
         \label{fig:hondasnedy}
   \end{figure}

In the Galactic halo, a few stars showing strong signatures of the weak $r$-process (or weak $s$-process) have been found and studied in detail. Two of those stars, HD~88609 ($\text{[Fe/H]}=-3.07$) and HD~122563 ($\text{[Fe/H]}=-2.77$) from \citet{Honda2007} are compared to ET0097 in Fig. \ref{fig:hondasnedy}. To ensure a useful comparison of the abundance pattern of the $n$-capture elements in these stars, they have all been normalized to the [Y/H] value of HD~88609. The absolute values of [X/H] for the $n$-capture elements are much higher in ET0097 than in the other, more metal-poor stars. In fact, when comparing to Fe, ET0097 is more enriched in Sr, Y, and Zr than the other two stars ([Y/Fe]$_{\text{ET0097}}=0.35$, [Y/Fe]$_{\text{HD~88609}}=-0.12$ and [Y/Fe]$_{\text{HD~122563}}=-0.37$). Also included in Fig.~\ref{fig:hondasnedy} (with the same normalization) is the $n$-capture rich star CS~22892-052 from \citet{Sneden2003}, which is believed to show a pure signature of the main $r$-process.

The relative abundance pattern in the $n$-capture elements of ET0097 is comparable to the two Honda stars (see Fig. \ref{fig:hondasnedy}), making it very likely that these stars were polluted by similar processes. The only exception is Pr, which seems to be much lower in ET0097 than in the others.\footnote{The Pr lines at 4179.4 and 4189.5~\text{\AA}, used by \citet{Honda2007}, are severely blended with CH molecular lines in ET0097, and therefore cannot be used for the abundance determination, but are consistent with no lines being observed. Instead the upper limit for Pr was determined from lines at 4408.8 and 4496.5~\text{\AA}, both giving similar results.} Apart from this one element, the abundance patterns of the three stars are comparable. The main $r$-process rich star, CS~22892-052, shows a completely different pattern, and so it is clear that the origin of its $n$-capture elements is different from the other stars. ET0097 and the two Honda stars show similar signatures of the weak $r$-process, possibly with some contamination from the main $r$-process.

Finally, it should be noted that the weak $s$-process in fast-rotating, zero-metallicity massive stars has been proposed to simultaneously enrich gas with C, N O and the lighter neutron-capture elements \citep{Chiappini2006,Frischknecht2012,Cescutti2014}. However, these models predict an enhancement of O comparable with the C enhancement, which is not observed in ET0097. It should also be noted that neither of the Honda stars are enhanced in carbon, [C/Fe]$\lesssim-0.40$ for both stars \citep{Honda2004b}, which is consistent with the idea that the enhancements of carbon and the light $n$-capture elements come from two different processes. This is also supported by the top panel of Fig. \ref{fig:light}, where the CEMP-no stars do not show any obvious trend of [Sr/Ba] with [Fe/H], different from the normal population, and the same is true for [Sr/Fe] with [Fe/H] (See, e.g., \citealt{Cescutti2014}, their Fig.~1).

\section{Possible formation scenario}\label{sec:PFS}

To explain the abundance pattern seen in ET0097, the formation scenario must include a plausible explanation for carbon and lighter neutron-capture enhancements at such high $\text{[Fe/H]}=-2.03\pm0.10$. Usually these are seen at lower values, $\text{[Fe/H]}\lesssim-3$, in the Galactic halo. The star ET0097 is also depleted in the heavier $n$-capture elements compared to other stars of similar iron abundance in Sculptor, see Fig.~\ref{fig:ele}. 

These peculiarities of ET0097 therefore seem to indicate that it was not formed from the same material as most other stars observed in Sculptor. One possible scenario is that this star was formed in one of the progenitor (mini)-halos of Sculptor that formed at high-redshift and initially evolved independently (e.g., \citealt{SalvadoriFerrara2009}), also sometimes called inhomogeneous mixing. 

Although the C-enhancement in CEMP-no stars is usually associated with faint SNe, which have relatively high C yields compared to their Fe-peak elements production (e.g., \citealt{Iwamoto2005}), pollution by faint SNe alone cannot enrich a gas cloud up to a metallicity of $\text{[Fe/H]}\approx-2$ \citep{Salvadori2012,CookeMadau2014}, and the general abundance pattern of the yields of such stars is very different from ET0097. However, as discussed in the previous section, it is possible that ET0097 was formed out of material that contained a mixture of yields from faint SNe and normal core collapse SNe. 

It has been shown that massive zero-metallicity SN are able to pollute small self-enriched systems up to high $\text{[Fe/H]}\approx-2$ \citep{Salvadori2007,Karlsson2008}. So one possibility is that ET0097 was formed in a mini-halo that had only been enriched by the first generation of stars, a population containing both zero-metallicity core collapse SN and faint SN. If that is the case, we should be able to estimate the required relative contributions of faint and "normal" primordial supernovae to simultaneously account for the observed C and Fe abundances and the lighter $n$-capture elements.

To test this idea, we assume a very simple scenario where this primordial population forms in a single burst of mass $M_\star$, with a Salpeter IMF in the mass range 10-100~M$_\odot$. A fraction $F_{F}$ of the star-forming gas goes into faint SN (as described by \citealt{Iwamoto2005} for HE0107-5240), and a fraction $F_{N}=1-F_{F}$ goes into normal zero-metallicity SN, (as described by \citealt{HegerWoosley2010} with standard mixing and energy $E_{SN}=1.2 \cdot 10^{51}$~erg)\footnote{In the normal SN case, the yields for carbon are taken from Table 8 in \citet{HegerWoosley2010}, and the iron taken from $^{56}$Ni yields in their Table 6 for the same mass values. Both elements are integrated in the same way over a Salpeter IMF.}.

The iron abundance of a gas cloud with mass $M_g$, and a mass of iron $M_\text{Fe}$ can be approximated by:
\begin{equation}\label{eq:FeH}
\text{[Fe/H]}=\log\left(\frac{ M_\text{Fe}}{M_g}\right)-\log\left(\frac{M_\text{Fe}}{M_\text{H}} \right)_\odot 
\end{equation}
Similarly, we get an expression for the mass of carbon in the gas, $M_\text{C}$,

\begin{equation} \label{eq:CFe}
\text{[C/Fe]}=\log\left(\frac{ M_\text{C}}{M_\text{Fe}}\right)-\log\left(\frac{M_\text{C}}{M_\text{Fe}} \right)_\odot 
\end{equation}

By assuming solar abundances from \citet{GrevesseSauval1998} the mass of Fe and C needed to enrich the gas up to [Fe/H]$=-2.03$ and [C/Fe]=0.8 is
\begin{eqnarray}
M_\text{Fe}&=&10^{-4.78}\cdot M_g \label{eq:Mfe}\\
M_\text{C}&=&14.2 \cdot M_\text{Fe} \label{eq:McMfe}
\end{eqnarray}

The total amount of iron and carbon produced in a star forming episode of total mass $M_\star$, is therefore 
\begin{eqnarray}
M_\text{Fe}&=&\mathcal{Y}_{N}(\text{Fe}) F_{N} M_\star + \mathcal{Y}_{F}(\text{Fe})F_{F} M_\star \label{eq:MFe2}\\
M_\text{C}&=&\mathcal{Y}_{N}(\text{C}) F_{N} M_\star + \mathcal{Y}_{F}(\text{C}) F_{F} M_\star \label{eq:MC2}
\end{eqnarray}
where the yields $\mathcal{Y}$ of Fe and C, are the masses of these elements (in M$_\odot$) that are produced and released into the environment by each M$_\odot$ of gas transformed into stars. The Fe yields in faint SNe are negligible, $\mathcal{Y}_{F}(\text{Fe})\ll \mathcal{Y}_{N}(\text{Fe})=2.0\cdot 10^{-3}$, and the carbon yields of the faint and normal SNe are respectively: $\mathcal{Y}_{F}(\text{C})=4.2\cdot 10^{-3}$ \citep{Iwamoto2005}, and $\mathcal{Y}_{N}(\text{C})=1.0\cdot 10^{-2}$ \citep{HegerWoosley2010}. Thus Eq. (\ref{eq:Mfe})-(\ref{eq:MC2}) give $F_{N}=0.19$ and $F_{F}=0.81$. 

By defining the star formation efficiency, $f_\star$, as the fraction of gas turned into stars, the total mass of the primordial population is
\begin{equation}
M_\star=f_\star M_g 
\end{equation} 
Combining this with Eq. (\ref{eq:Mfe}) and (\ref{eq:MFe2}), gives $f_\star$=0.044. This is an upper limit, since more realistic calculations should take into account that stars with different masses do not explode all at once and some of the gas is ejected by SNe of massive stars before the lower mass stars start to contribute, leaving less gas to enrich.

This scenario is also able to explain the overabundance of the lighter $n$-capture elements (Sr, Y, Zr), seen in ET0097, if certain requirements are fulfilled. \citet{ArconesMontes2011} describe the weak $r$-process in neutrino-driven winds from core collapse supernovae of progenitor mass 10~M$_\odot \leq M \leq$~25~M$_\odot$. Using their yields and going through similar calculations to those above, the weak process has to occur in $\sim$10\% of the total stellar mass formed as a normal SN in the mass range 10-25~M$_\odot$, to account for the Sr, Y, and Zr abundances observed in ET0097. 

\citet{HegerWoosley2010} have shown that their yields are consistent with the general abundance pattern seen in low-metallicity halo stars (such as in \citealt{Cayrel2004}) for other elements up to Zn, and it is therefore reasonable to conclude that they are also consistent with ET0097. 
 
With this very simple calculation, we show that ET0097 is compatible with having been enriched with the first stellar generation. We are able to explain both the high iron abundance, and the overabundances of carbon and the lighter $n$-capture elements.

Another (similar) possibility is that this star was formed in a (mini)-halo, where the first stellar generation was dominated by faint SN, and then normal (nonzero-metallicity) Type II SNe \citep{WoosleyWeaver1995}, which have comparable C and Fe yields to the zero-metallicity case, enriched the gas up to $\text{[Fe/H]}\approx-2$.

In both of these scenarios, it is necessary that the (mini)-halo is large enough to retain some of its gas for the next generation(s) of stars, and that the stellar population of zero-metallicity stars is dominated by faint SN. This is consistent with \citet{deBennassuti2014} who show that a stellar population of zero-metallicity stars dominated by faint SN is able to produce the CEMP fraction observed in the halo. 

\section{The CEMP-no fraction in Sculptor}
In total, 22 stars in Sculptor with $\text{[Fe/H]}\leq-2$ have C measurements or upper limits (\citealt{Frebel2010,Tafelmeyer2010,Kirby2012,Starkenburg2013}; and this work). Only one of them falls into the category of a CEMP-no star, making the fraction $4.5^{+10.5}_{-3.8}$\% (errors are derived using \citealt{Gehrels1986}).

In the Galactic halo, the proportion of CEMP-no stars (when the RGB stars have been corrected for internal mixing and CEMP-$s$ and CEMP-$s/r$ stars have been excluded from the sample) is $20\pm2$\% (compilation of data and correction for mixing comes from \citealt{Placco2014}\footnote{The \citet{Placco2014} sample is based on the most recent version of the SAGA database \citep{Suda2008} and the compilation of literature data by \citet{Frebel2010} and data published since then. For individual references see \citet{Placco2014}.}, with Poisson errors derived from \citealt{Gehrels1986}). If this fraction was the same in Sculptor, $p=0.20$, then from a sample of $N=22$ stars, the expected number of CEMP-no stars with $\text{[C/Fe]}\geq0.7$ is $R_{exp}=Np\pm\sqrt{N(1-p)p}=4.4\pm 1.9$. However, only one is found. In the lower metallicity range, $\text{[Fe/H]}\leq-3$, the fraction of CEMP stars in the Galactic halo is $p=0.43^{+0.06}_{-0.05}$ \citep{Placco2014}. In Sculptor, $N=8$ stars in this range have measured carbon or upper limits, so the expected number of stars with $\text{[C/Fe]}\geq 0.7$ (once they have been corrected for mixing), is $R_{exp}=3.4\pm 1.4$, but none are found. The fraction of CEMP-no stars in the currently observed Sculptor sample is thus lower than in the Galactic halo, and the difference is statistically significant.

This ratio can be affected by the fact that no stars with $\text{[Fe/H]}\leq-4.0$ have been found so far in Sculptor, while those stars predominantly fall into the CEMP-no category in the halo. Looking at the range $-2.5\leq\text{[Fe/H]}\leq-2$, the fraction of CEMP-no stars in Sculptor is  $8^{+19}_{-7}$\%, which is consistent with that found in the halo, $5^{+3}_{-2}\%$ (data coming from \citealt{Placco2014}, with Poisson errors derived from \citealt{Gehrels1986}). The expected number of CEMP-no stars in this range, should the Sculptor fraction be the same as in the halo, is $R_{exp}=0.6^{+0.8}_{-0.6}$ stars, which is consistent with the one star found. Only one observed (C-normal) star in Sculptor falls in the range $-3<\text{[Fe/H]}<-2.5$, so little can be said about the CEMP-no fraction there. However, for the lowest metallicity range, $-4\leq\text{[Fe/H]}\leq-3$, no CEMP-no star is found in Sculptor out of a sample of eight stars, giving the Poisson upper limit of the fraction, $23\%$. In the same metallicity range in the halo, the fraction is $39^{+6}_{-5}\%$. The expected number of CEMP-no stars in the Sculptor sample, should the fraction be the same as in the halo is $R_{exp}=3.1\pm1.4$ stars, while none are found. Although still poorly constrained due to low number statistics, the CEMP-no fraction in Sculptor is therefore consistent with the Galactic halo at higher metallicities ($-2.5\leq\text{[Fe/H]}\leq-2$), while it appears to be different at the lowest metallicity end, $-4\leq\text{[Fe/H]}\leq-3$ (see also: \citealt{Starkenburg2013}).

It remains puzzling that no CEMP-no stars are found at lower metallicities in Sculptor, while ET0097 has a rather high metallicity for such objects, $\text{[Fe/H]}=-2.03\pm0.10$. If the CEMP-no stars indeed show imprints of the very first stars, they would be expected to be more common at the lowest metallicities, also in dwarf spheroidal galaxies such as Sculptor. This apparent discrepancy is not easily explained, and it cannot be excluded that CEMP-no stars are a mixed population with different formation scenarios (an overview is given in \citealt{Norris2013}). However, we want to emphasize that the CEMP-no fraction in Sculptor is still very poorly constrained, with an upper limit of $\sim$25\% both at the high and the low-metallicity end, and not very constraining lower limits ($<2\%$). Although the CEMP-no fraction at the low-metallicity end in Sculptor seems to be different from the halo, a similar trend is still possible where the fraction increases with lower [Fe/H], and should be expected. The effect of the environment on the CEMP-no fraction is still not well understood, and will be explored in greater detail in Salvadori et al. in prep.


%
\section{Conclusions}
After unusually strong CN molecular lines were discovered in the star ET0097, a follow-up spectrum at high-resolution and over a long wavelength range was taken with the ESO/VLT/UVES spectrograph. Detailed abundance analysis shows that with [C/Fe]=0.51$\pm$0.10, ET0097 is the most carbon-rich VMP star known in Sculptor. Having a luminosity of $\log L_\star/L_\odot=3.1$, this star is expected to have undergone mixing, lowering the carbon at the surface of the star and increasing the nitrogen. This is confirmed by the high N of the star, [N/Fe]=1.18$\pm$0.20, the low isotope ratio $\log^{12}$C/$^{13}$C=0.77$\pm$0.03, and the low Li abundance $\log \epsilon(\text{Li})<0.17$. The original C abundance of ET0097 is therefore estimated to be $\text{[C/Fe]}\approx0.8$, making this the only known CEMP in Sculptor.

The star shows normal abundances for all alpha and Fe-peak elements from O to Zn, consistent with what is seen both in the Galactic halo and Sculptor for giants of similar metallicities. Compared to other stars in Sculptor, ET0097 is enhanced in the lighter $n$-capture elements (Sr, Y, Zr), and shows low abundances of the heavier $n$-capture elements, making this a CEMP-no star. The abundance ratios [Sr,Y,Zr/Ba] are high, especially in relation to the Fe abundance of the star. The abundance pattern of the $n$-capture elements is comparable to what is seen in stars that are believed to have been polluted by the weak $r$-process.

One possible scenario to explain the peculiar abundance pattern of ET0097 is that the star formed in one of the low-mass progenitor halos of Sculptor, which had only been enriched by a primordial population consisting of a mixture of faint SN ($\sim$80\%) and zero-metallicity core collapse SN ($\sim$20\%). Another possibility is that this star was formed in a halo where the first stellar generation consisted of faint SN only, and then normal Type~II SNe polluted the gas up to $\text{[Fe/H]}\approx-2$.

In addition to the abundance analysis for ET0097, ESO/VLT/FLAMES data in the wavelength range 9100-9250~\text{\AA} was used to determine estimates and upper limits for carbon in 85 Sculptor stars, including 11 stars with $\text{[Fe/H]}\leq-2$. No other star in the sample was found to be carbon-enhanced.

In the Galactic halo, \citet{Placco2014} have carefully determined the fraction of CEMP stars in a sample of $\sim$500 stars with $\text{[Fe/H]}\leq-2$. The RGB stars in their sample have been corrected for internal mixing, and CEMP-$s$ and CEMP-$s/r$ stars have been excluded. Using these very detailed results, we are able to compare the CEMP-no fraction observed in Sculptor to that seen in the halo. The fraction of CEMP-no to C-normal stars in the entire sample of observed Sculptor stars with $\text{[Fe/H]}\leq-2$, is $4.5^{+10.5}_{-3.8}$\%. This is lower than that seen in the Galactic halo, $20\pm2\%$ \citep{Placco2014}, and the difference is statistically significant.

If we explore this further in different metallicity bins, then for $-2.5\leq\text{[Fe/H]}\leq-2$, the observed CEMP-no fraction in Sculptor is  $8^{+19}_{-7}$\%, consistent with that found in the halo, $5^{+3}_{-2}\%$. At the lowest metallicity end in Sculptor, $-4\leq\text{[Fe/H]}\leq-3$, the CEMP-no fraction is $0^{+23}$\%, which is statistically significantly lower then that observed in the Galactic halo, $39^{+6}_{-5}\%$. The carbon measurements in Sculptor are still few, so the CEMP-no fraction is poorly constrained, but, at least at the lowest metallicity end, the CEMP-no fraction in Sculptor seems to be fundamentally different from what is seen in the Galactic halo.

\begin{acknowledgements}

We thank ESO for granting us Directors Discretionary time to allow us to rapidly follow up this very interesting star. The authors are indebted to the International Space Science Institute (ISSI), Bern, Switzerland, for supporting and funding the international team "First stars in dwarf galaxies". \'{A}.~S. thanks Anna Frebel for useful advice and insightful suggestions. S.~S. acknowledges support from the Netherlands Organisation for Scientific Research (NWO), VENI grant 639.041.233. E.~S. gratefully acknowledges the Canadian Institute for Advanced Research (CIFAR) Global Scholar Academy.
\end{acknowledgements}
%



\end{document}